\documentclass[traditabstract]{aa} 
\usepackage{graphicx}
\usepackage{txfonts}
 \usepackage[T1]{fontenc}
\usepackage{color}
\usepackage[]{natbib}
\usepackage[switch]{lineno}
\usepackage[utf8]{inputenc}
\usepackage{amsmath}

	\newcommand{\arc}{$^{\prime\prime}$}
	\newcommand{\msun}{\ensuremath{\mathrm{M}_{\odot}}}
	\newcommand{\lsun}{L$_{\odot}$}
	\newcommand{\sfr}{M$_{\odot}$ yr$^{-1}$}

	\newcommand{\hb}{${\rm H\beta}$}
	
	\newcommand{\lagn}{${\rm L_{AGN} }$}
	\newcommand{\mom}{$\dot{P}$} 
	
	\newcommand{\mion}{$\dot M_{o}$}

	\newcommand{\kms}{${\rm km/s}$}

	\newcommand{\oiii}{[O{\sc iii}]$\lambda$5007}

\begin{document}
   \title{Ionised outflows in $z\sim 2.4$ quasar host galaxies \thanks{based on Observations collected  at the European Organisation for Astronomical Research in the Southern Hemisphere, Chile, P.ID: 086.B-0579(A)  }}

   \author{S. Carniani \inst{1,2,3,4},
          A. Marconi  \inst{1,2},
          R. Maiolino \inst{3,4},
          B. Balmaverde \inst{2},
          M. Brusa \inst{5,6},
          M. Cano-D\'iaz \inst{7}
          C. Cicone   \inst{3,4,8},
          A. Comastri \inst{6},
          G. Cresci  \inst{1,2},
          F. Fiore \inst{9},
      	  C. Feruglio \inst{9,10,11},
          F. La Franca \inst{12},
          V. Mainieri \inst{13},
          F. Mannucci \inst{2},
          T. Nagao \inst{14},
          H. Netzer \inst{15},
          E. Piconcelli \inst{9},
          G. Risaliti \inst{2},
          R. Schneider \inst{9},
	  O. Shemmer \inst{16}
                 }

   \institute{Dipartimento di Fisica e Astronomia, Universit\`a di Firenze, Via G. Sansone 1, I-50019, Sesto Fiorentino (Firenze), Italy 
                      \and
             INAF - Osservatorio Astrofisico di Arcetri, Largo E. Fermi 5, I-50125, Firenze, Italy 
                      \and
	Cavendish Laboratory, University of Cambridge, 19 J. J. Thomson Ave., Cambridge CB3 0HE, UK 
		\and
	Kavli Institute for Cosmology, University of Cambridge, Madingley Road, Cambridge CB3 0HA, UK 
		\and
	Dipartimento di Fisica e Astronomia, Universit\`a di Bologna, viale Berti Pichat 6/2, 40127 Bologna, Italy  
		\and 
	INAF - Osservatorio Astronomico di Bologna, via Ranzani 1, 40127 Bologna, Italy 
	\and
	Instituto de Astronom\'ia - UNAM, Mexico City, Mexico 
		\and
	 Institute for Astronomy, Department of Physics, ETH Zurich, Wolfgang-Pauli-Strasse 27, CH-8093 Zurich, Switzerland 
	\and
	INAF - Osservatorio Astronomico di Roma, via Frascati 33, 00040 Monteporzio Catone, Italy 
	\and
	Scuola Normale Superiore, Piazza dei Cavalieri 7, 56126 Pisa, Italy
	\and 
 IRAM - Institut de Radio Astronomie Millime\'trique, 300 rue de la Piscine, 38406 Saint Martin d'He\'res, France 
 	 \and
	 Dipartimento di Matematica e Fisica, Universit\`a Roma Tre, via della Vasca Navale 84, I-00146 Roma, Italy 
	\and 
	European Southern Observatory, Karl-Schwarzschild-str. 2, 85748 Garching bei M\"unchen, Germany 
	\and
	Research Center for Space and Cosmic Evolution, Ehime University, Bunkyo-cho 2-5, Matsuyama, 790-8577 Ehime, Japan 
	\and
	School of Physics and Astronomy, The Sackler Faculty of Exact Sciences, Tel-Aviv University, Tel-Aviv 69978, Israel 
	\and
	Department of Physics, University of North Texas, Denton, TX 76203, USA
		  }

  \abstract{   
\emph{Aims}. AGN-driven outflows are invoked by galaxy evolutionary models to quench star formation and to explain the origin of the relations observed locally between super massive black holes and their host galaxies. This work aims to detect the presence of extended ionised outflows in  luminous quasars where we expect the maximum activity both in star formation and in black hole accretion. Currently, there are only a few studies based on spatially resolved observations of outflows at high redshift, $z>2$.\\
 \emph{Methods}. We analyse a sample of six  luminous (${\rm L>10^{47} \ erg/s}$) quasars at $z\sim2.4$, observed in H-band using the near-IR integral field spectrometer SINFONI at VLT. 
 We perform a kinematic analysis of the  [O{\sc iii}] emission line at $\lambda = 5007\AA$.  \\
 \emph{Results}. We detect fast, spatially extended outflows in five out of six targets. 
 \oiii\  has a complex gas kinematic, with blue-shifted velocities of a few hundreds of \kms\ and line widths up to 1500 \kms. Using the spectroastrometric method we infer size of the ionised outflows of up to $\sim$2 kpc.  The properties of the ionised outflows, mass outflow  rate, momentum  rate and kinetic power, are correlated  with the AGN luminosity. The increase in outflow rate with increasing AGN luminosity is consistent with the idea that a luminous AGN  pushes away the surrounding gas through  fast outflows driven by radiation pressure,  which depends on the  emitted luminosity. \\
\emph{Conclusions}. We derive  mass outflow rates of about 6-700 \sfr\ for our sample, which are lower than those observed in molecular outflows. 
Indeed physical properties of ionised outflows show dependences on AGN luminosity which are similar to those of molecular outflows but indicating that the mass of ionised gas is smaller than that of the molecular one.
Alternatively, this discrepancy between ionised and molecular outflows could be explained with different acceleration mechanisms.
  }

\authorrunning{Carniani et al.}
\titlerunning{Ionised outflows in $z\sim 2.4$ quasar host galaxies}
 \maketitle

\section{Introduction}\label{sec:introduction}

Feedback mechanisms from quasars (QSOs)  are considered to be  crucial  for galaxy evolution (see \citealt{King:2010}, \citealt{Fabian:2012} and \citealt{King:2015}). During the  bright active phase, Active Galactic Nuclei (AGN) are believed to drive energetic outflows that expel gas at  large scales from their  host galaxies (e.g \citealt{Granato:2004}, \citealt{Di-Matteo:2005}, \citealt{Menci:2008},  \citealt{King:2010}, \citealt{Zubovas:2012}, \citealt{Fabian:2012}, \citealt{Faucher-Giguere:2012}, \citealt{Zubovas:2014}, \citealt{Nayakshin:2014}, \citealt{Costa:2014}, \citealt{Costa:2015}), hence removing the supply of cold gas required for star formation (SF) activity. 
According to some of these models, the black hole (BH) achieves a critical mass during the later stages of a merging event  and the energy output of the associated AGN  is so large that  radiation pressure drives a fast outflow in the nuclear region which  sweeps away the  gas in the host galaxy.
When the feedback phase is over, the stellar population in the host galaxy 
continues to grow mainly through minor or major mergers  with  other galaxies. (e.g. \citealt{Baldry:2004}, \citealt{Perez-Gonzalez:2008a}). At  the same time the lack of gas around the nucleus slows down the growth of the BH. 
Models invoke  feedback mechanisms to explain the origin of the correlation between the mass of  supermassive BHs and the mass and velocity dispersion of host galaxy bulges observed in the local Universe (e.g. \citealt{Magorrian:1998, Marconi:2003, Ferrarese:2005, Kormendy:2013}). 
Essentially, the observed correlations can be explained as the result of the balance between the outward radiation force generated by the AGN and the inward gravitional force of the host galaxy (e.g. \citealt{King:2010, Fabian:2012}).

Quasar-driven outflows extending to kpc-scales have been resolved both locally  (e.g \citealt{Feruglio:2010, Rupke:2011,Cicone:2012, Rupke:2013,Rodriguez-Zaurin:2013, Feruglio:2013, Feruglio:2013a, Cicone:2014,Rodriguez-Zaurin:2014,Aalto:2015, Feruglio:2015}) and at high redshift (e.g. \citealt{Alexander:2010,Nesvadba:2011,Harrison:2012,Maiolino:2012,Cano-Diaz:2012,Harrison:2014,Cresci:2015, Cicone:2015}) in ionised, atomic and molecular gas. 
Despite big advances in data quality and analysis in the past decade, the main  properties of these energetic outflows  remain largely unknown. 
The exact outflow morphologies and driving mechanisms are still poorly known: it is  debated whether their morphology is conical or shell-like, and it is also unclear what is the physical process responsible for the coupling of the energy/momentum released by the central AGN with the galaxy interstellar medium (e.g. inner winds and shocks, radiation pressure on dust). According to one of such scenarios, 
the fast wind,  accelerated close to the BH by radiation pressure, shocks the ISM of the host galaxy creating a bubble of hot gas which expands at large velocities ($\sim 1000$ km/s). If the post shock material does not cool efficiently, energy is conserved, the bubble expands adiabatically and the outflow is energy-driven. On the other hand, if the post shock material cools efficiently, due for example to Compton cooling by AGN photons, only  momentum is conserved and the outflow is momentum-driven. Since the efficiency of the cooling process, dominated by inverse Compton scattering, drops with increasing shock radius, there is a critical distance ($\sim$ 100 pc - 1 kpc) beyond which the extended outflows can  only be  energy-driven   \citep{Zubovas:2012}.
 Indeed, recent  observations of  kpc-scale outflows in local AGNs support this scenario (\citealt{Cicone:2014}, \citealt{Feruglio:2015},  \citealt{Tombesi:2015}).

Moreover it is not clear whether molecular and ionised outflows are 
accelerated by the same mechanism, whether they have the same spatial distribution or whether they occur on similar timescales. So far, there are only a few observations of both molecular and ionised AGN-driven outflows in the same galaxy: SDSS J1356+1026 is an example of an obscured QSO where  molecular and ionised outflows have different properties, i.e. outflow rates, velocities, radii, morphologies  and time scales (\citealt{Greene:2012, Sun:2014}).
The forbidden  emission line doublet [O{\sc iiii}] at $\lambda=5007,4959\AA$ is a good tracer of  ionised outflows on large scales since it can not be produced at  high densities and so it cannot trace the sub-parsec scales of the Broad Line Region (BLR). In the presence of outflows, the spectral profile of the \oiii\ emission line can be highly asymmetric, with a broad, blue-shifted wing that is  rarely  observed in star-forming regions.  For this reason, the \oiii\ emission line has been used to identify outflowing  ionised gas in low-redshift and high-$z$ AGNs.
Recent integral field observations of the \oiii\ emission  have provided  quantitative measurements of the outflowing 
gas properties in AGN at low redshifts $z < 0.5$ (e.g. \citealt{Greene:2011}; \citealt{Harrison:2014}). 
At higher redshift, $z\sim2$, \cite{Alexander:2010}   found evidence for galactic-scale ionised outflows by mapping the \oiii\ emission line in ULIRGs (Ultra Luminous IR Galaxies)  hosting type 1 AGNs. \cite{Cano-Diaz:2012} observed a ionised outflow, extended up to $\sim$8 kpc, in a QSO at $z\sim2.4$. Finally,  \cite{Brusa:2015}  and \cite{Perna:2015} detected  broad \oiii\ blue wings in a sample of obscured AGN revealing the presence of outflows extended over several kiloparsecs. In one case of the latter sample, the ionised outflow, extended up to $\sim12$ kpc, is anti-correlated with the presence of star formation in the host galaxy \citep{Cresci:2015}, similarly to what found by \cite{Cano-Diaz:2012}.




In this paper, we present a kinematical analysis of the \oiii\  emission line observed in six high luminous (${\rm L_{\rm bol}} > 10^{47}$ erg/s) quasars at $z\sim~2.4$. The line profiles  and the velocity maps obtained by the kinematical analysis show the presence of ionised outflows extended on  scales larger than $2-3$~kpc from the nucleus. This is one of the first observations of extended ionised outflows  in QSO or type 1 AGN; indeed
most  AGN-driven outflows have been observed so far in  type 2 AGN. 
We present a new method to measure outflow properties from IFU ( Integral-Field Unit) data when the source is only marginally spatially resolved. We show that the typical signature of outflows, i.e. asymmetric line profiles and blue-shifted components in the velocity maps, can be described by two point sources (the central AGN source  and the outflowing material) separated by a distance of a few kpc. We find that the ionised gas likely traces only a fraction of the total  outflowing gas, unless    ionised outflows have a different origin than the molecular ones. 

The paper is organised as follows: in Section 2 we present the sample selection and properties, in Section 3 we show the data analysis and spectral fitting. The results of our data analysis are presented in Section 4. Finally, in Section 5 we discuss  the main results, i.e. the nature of ionised outflows and the comparison with other observations. 
 A $H_0 = 67.3$ km s$^{-1}$ Mpc$^{-1}$, $\Omega_{M} = 0.315$,  $\Omega_{\Lambda} = 0.685$ cosmology is adopted throughout  this work \citep{Planck-Collaboration:2014aa}.

\section{Sample selection}\label{sec:data}

 We selected six QSOs at $z\sim$2.4 with large \oiii\
 equivalent widths (> 10$\AA$ in the rest frame) and bright H-band magnitude
 ($< 16.5$ mag) from the sample of \cite{Shemmer:2004}, \cite{Netzer:2004} and
 \cite{Marziani:2009}.  The luminosities of the objects in our sample 
 are in the range $ L= 10^{47}- 10^{48}$~erg/s, making them the highest luminosity sources where outflows have been spatially mapped.  
  These characteristics are chosen to maximise 
 our chances to detect signatures of feedback on the host galaxy. As
 explained  in Section \ref{sec:introduction}, the cosmic epoch   corresponding to $z\simeq2$ is crucial  for
  the growth  of the most massive galaxies and
 black holes, and shows  the maximum activity both in star
 formation and in BH accretion (e.g., \citealt{Madau:2014aa}). 
The selected objects,  at the high end  of the QSO luminosity function,
 are those where we expect to detect feedback in action, at least according to current galaxy evolutionary models (e.g. \citealt{Hopkins:2006}).
 Moreover, the targets
 were selected depending on their particular [O{\sc iii}]  profile.
 Indeed, the large \oiii\ equivalent width should allow us to easily recover the
 kinematical maps of the ionised gas, a challenging task in luminous
 QSOs where narrow emission lines are usually weak (e.g
 \citealt{Netzer:2004}).  Finally, the bright ionised  
line profile can be easily de-blended from the broad \hb\ and
 FeII emission, which are associated to the BLR.
 Indeed, based on near-IR spectra available from the literature \citep{Shemmer:2004, Netzer:2004, Marziani:2009}, we also chose our objects such that the contamination from broad FeII emission is expected to be small.
 
Basic properties of  the observed  objects are given in Table~\ref{tab:summary}.
One of our  sources, 2QZJ0028, is already studied by \cite{Cano-Diaz:2012}, and was reobserved by using much longer integration time, confirming their results.

\section{Observations and data reduction}\label{sec:data2}

The targets, whose properties are described in Table \ref{tab:summary},   were observed in February and September 2012 using the near-IR integral field
spectrometer SINFONI (Spectrograph for INtegral Field Observations in
the Near Infrared) at the VLT (Very Large Telescope). The observations were obtained in H-band ($\lambda \sim 1.45 - 1.85\ \mu$m, where \oiii,
[O{\sc iii}]4959 and \hb\ are detected), in seeing limited mode with 0.250\arc\ spatial scale and  medium spectral resolution  of R = 3000. In all observations, the object was moved across the 8\arc$\times$8\arc\ field of  view in order to perform the sky subtraction. A standard star for telluric correction and flux calibration was observed shortly after or before the on-source exposures. 
The total on-source
 integration time is over 3h per  target. The  airmass are different for each target and spanning a range between $\sim1.0$ and $\sim1.4$.

After removing cosmic rays  from the raw data using the L.A. Cosmic procedure (\citealt{van-Dokkum:2001}), we used the ESO-SINFONI pipeline to reduce the data. 
The final data cubes produced by the ESO-SINFONI
pipeline have a spatial scale of 0.125\arcsec$\times$0.125\arcsec\ and a
field of view of about 8\arc$\times$8\arc.  The estimated angular resolution is 
$\sim0.4-0.6$\arcsec, based on a 2D-Gauusian fitting of the flux map of the spatially unresolved broad \hb\ line (see Section \ref{sec:fitting}).


\begin{table*}
\caption{Properties of our quasar sample}           
\label{tab:summary}      
\centering          
\begin{tabular}{l l c c c c}    
			 \\ 
\hline
ID & Target Name &  $\lambda_{0}$ [$\mu$m]  &  $\Delta\lambda_{\rm rest}$ [\AA]  & Redshift & $\lambda L_{5100}$ [$10^{46}$ erg/s]  \\ 
\hline
LBQS0109 & LBQS 0109+0213 & $1.68\pm0.04$   & $28.5\pm0.2$     & $2.35\pm0.08$   & $4.9\pm0.8$    \\
2QZJ0028 & 2QZ J0028-2830$^{\rm (a)}$ &$1.70\pm0.04$   &$27.9\pm0.2$     &$2.40\pm0.09$   &$3.1\pm0.6$  \\
HB8905  & HB89 0504+030  &$1.75\pm0.05$   &$22.7\pm0.2$     &$2.48\pm0.09$   &$2.9\pm0.6$   \\
HE0109 & HE 0109-3518   &$1.706\pm0.003$ &$17.342\pm0.011$ &$2.407\pm0.007$ &$7.4\pm1.5$  \\
HB8903 & HB89 0329-385  &$1.720\pm0.013$ &$19.31\pm0.06$   &$2.44\pm0.03$   &$5\pm2$     \\
HE0251 &HE 0251-5550   &$1.68\pm0.03$   &$31.81\pm0.17$   &$2.35\pm0.05$   &$6.8\pm1.4$    \\
\hline	
\multicolumn{6}{l}{\tiny (a): this is  the same target of \cite{Cano-Diaz:2012} which has been re-observed.} \\
\end{tabular}
	
\end{table*}

\section{Data analysis}\label{sec:fitting}

\begin{figure*}
   \centering
\includegraphics[width =0.49\textwidth]{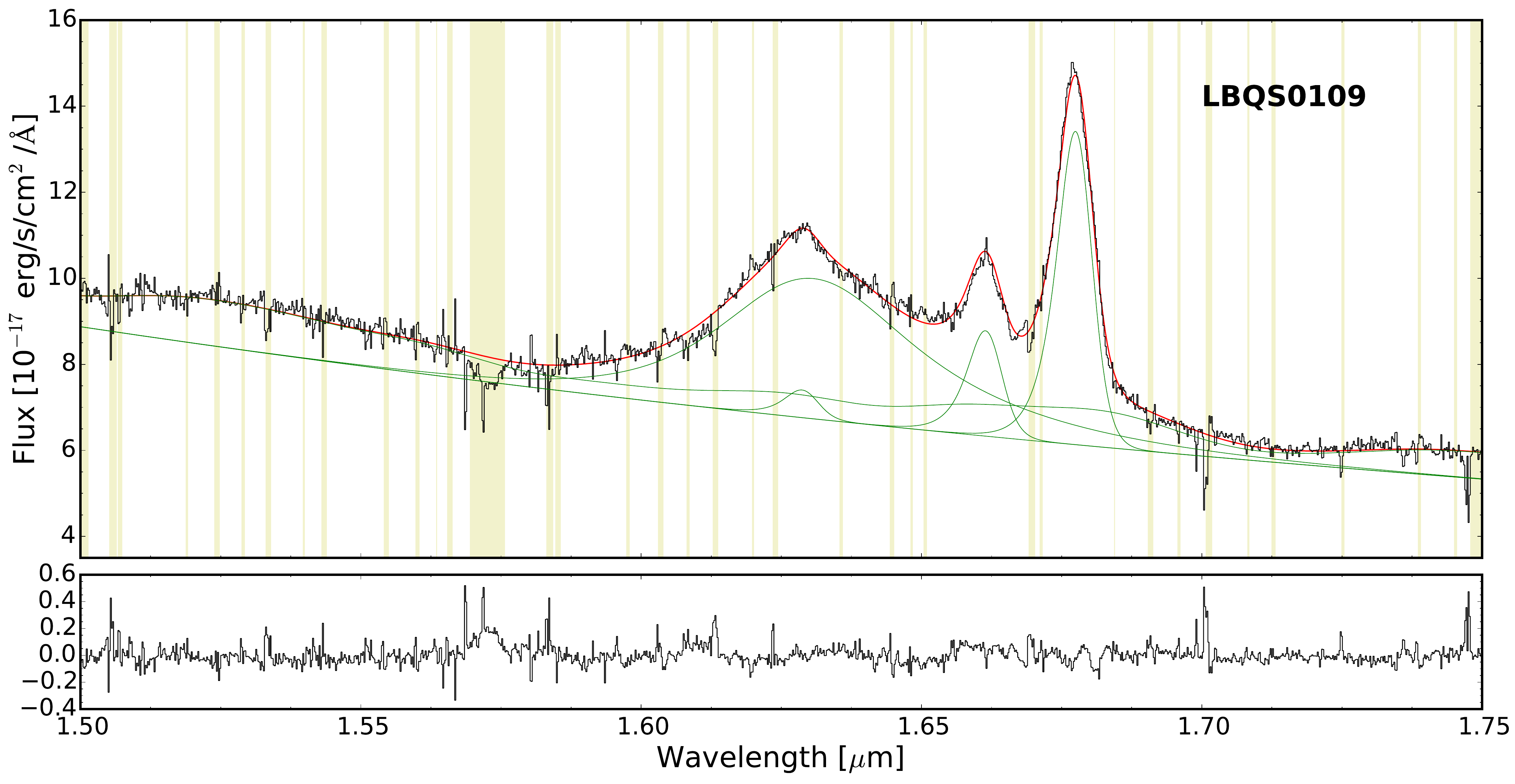}
\includegraphics[width =0.49\textwidth]{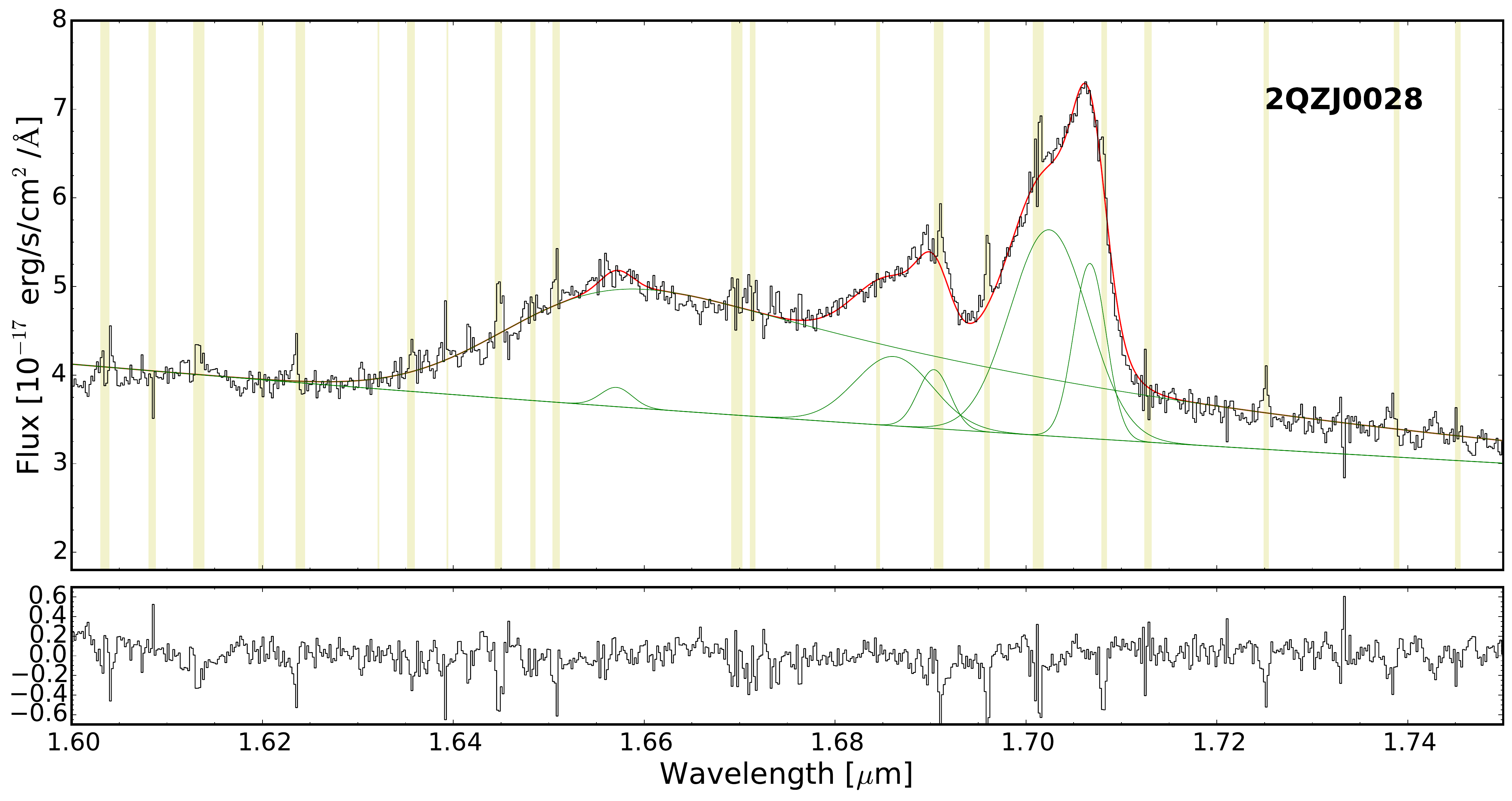}
\includegraphics[width =0.49\textwidth]{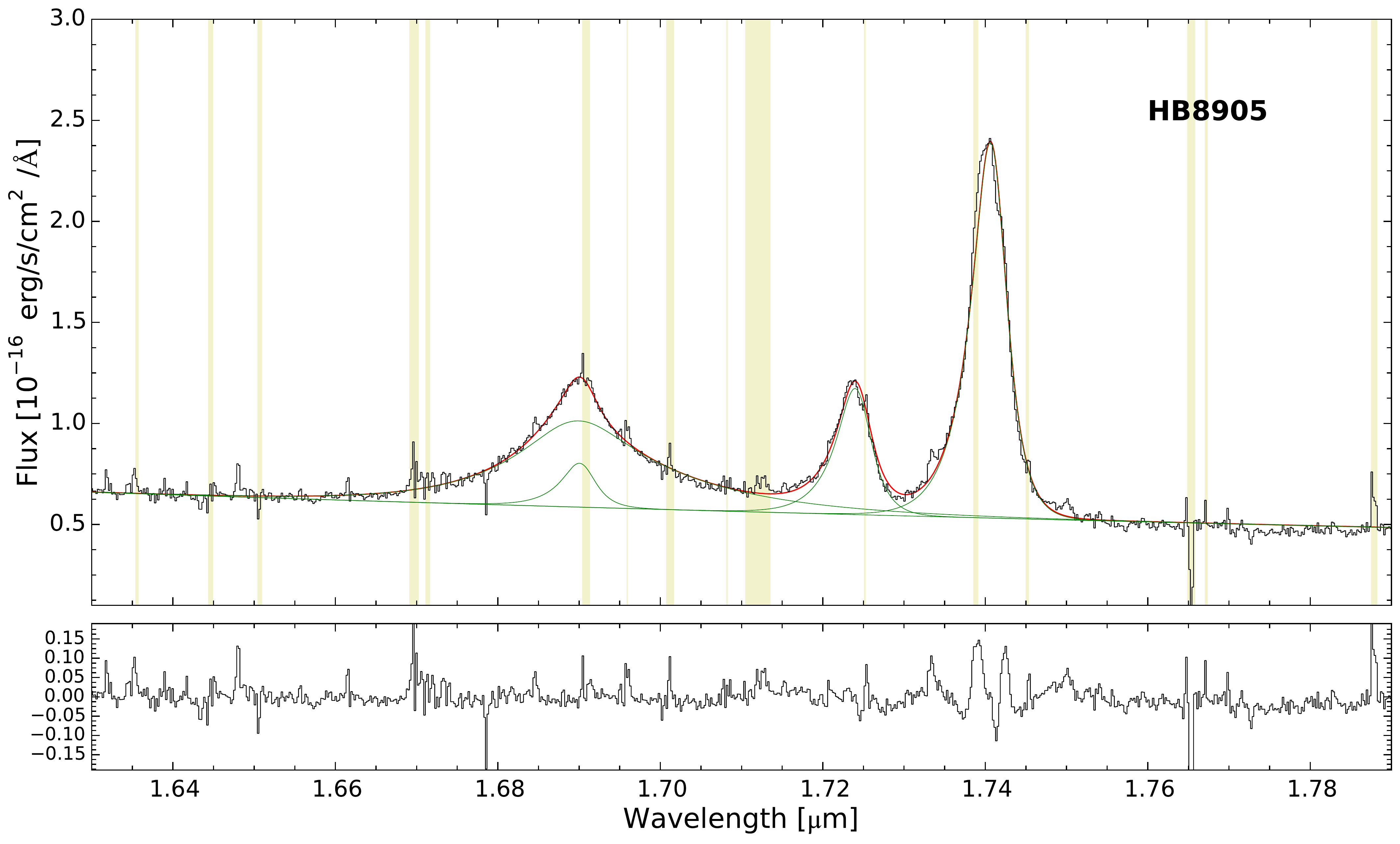}
\includegraphics[width =0.49\textwidth]{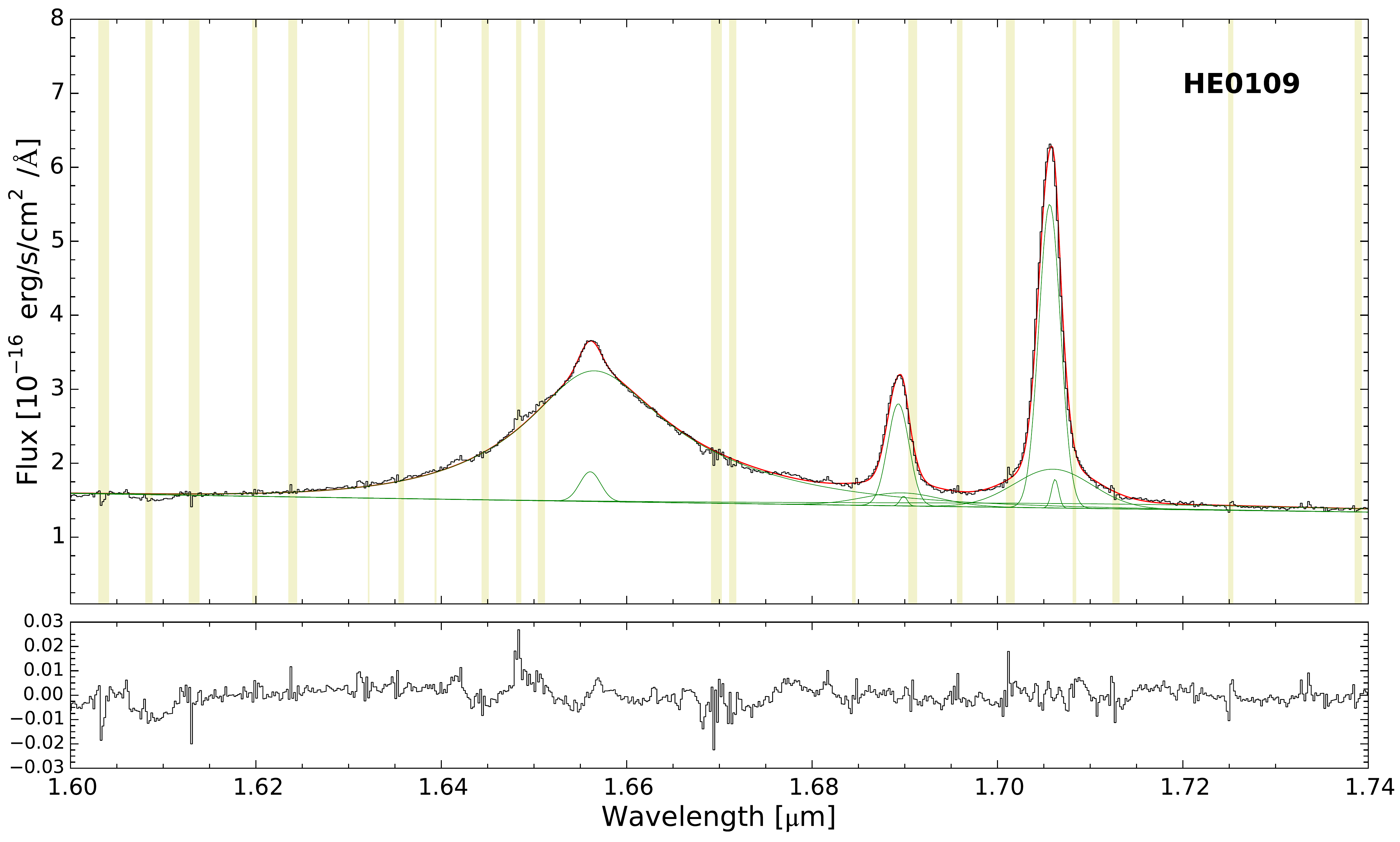}
\includegraphics[width =0.49\textwidth]{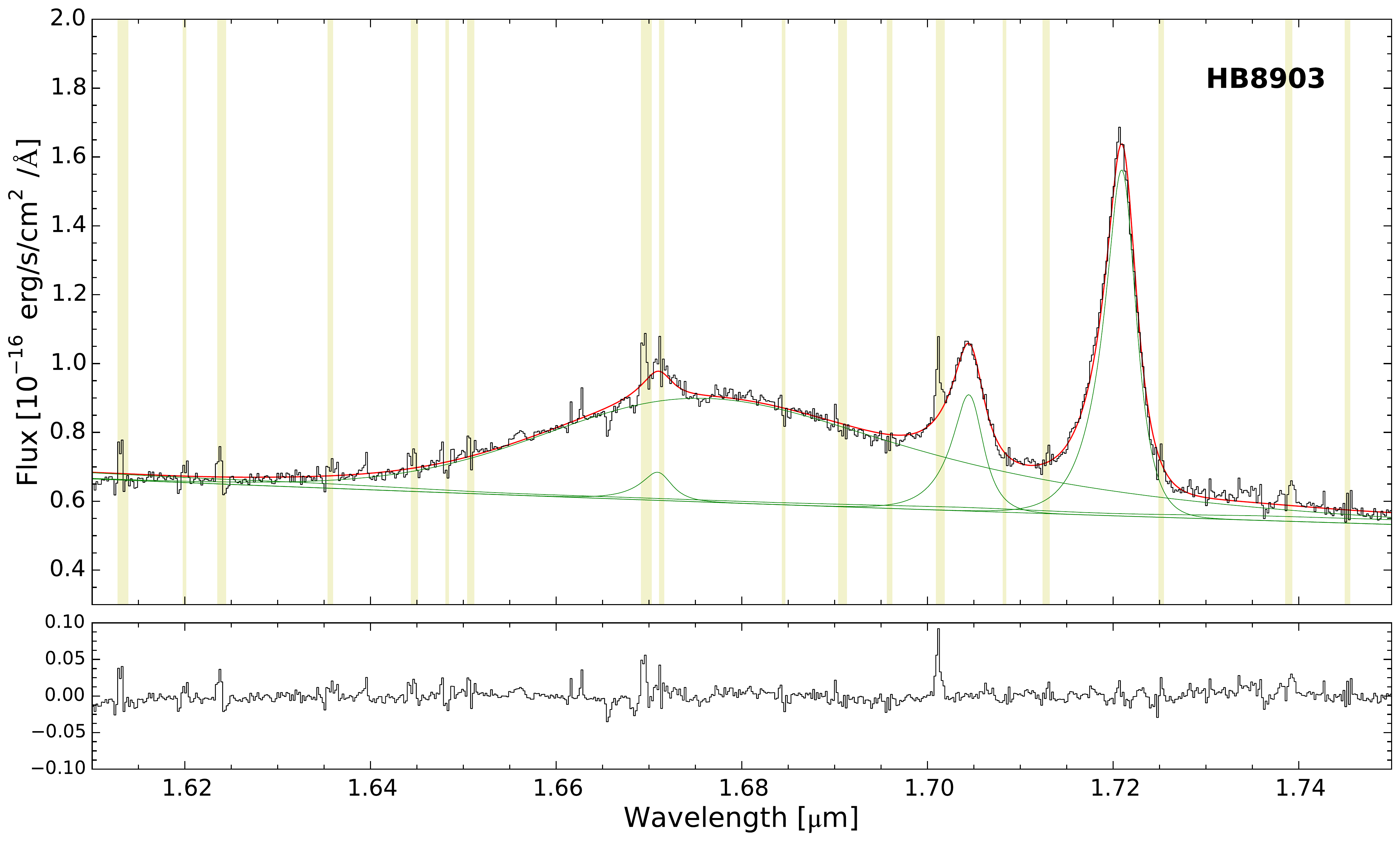}
\includegraphics[width =0.49\textwidth]{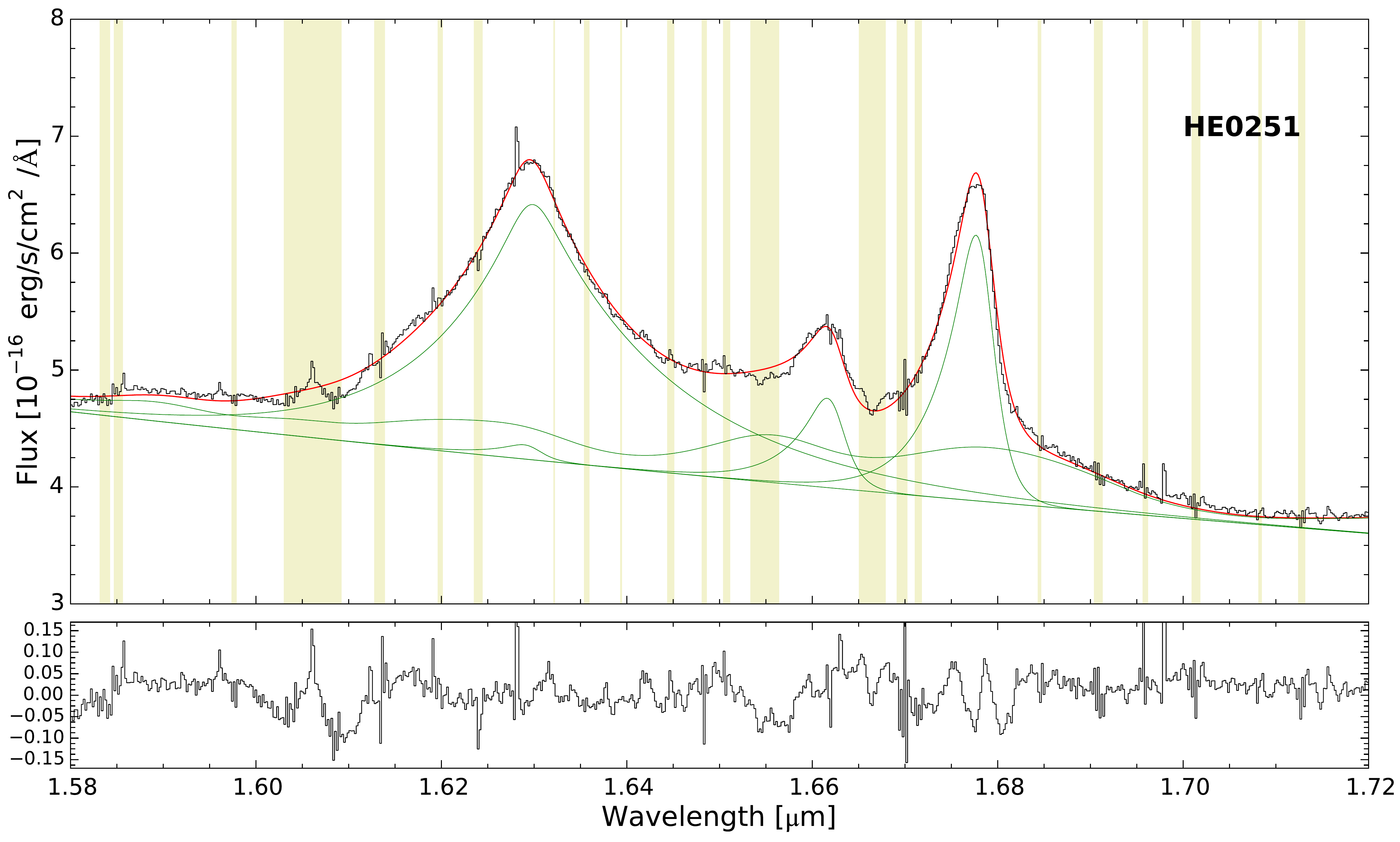}

\caption{ \emph{Upper panel}: The spectra of the six QSOs if our sample. Each spectrum is extracted from a nuclear region of 0.25\arc$\times$025\arc\ ($2\times2$ pixel). The different components in the fit for each line (\hb,[O{\sc iii}] and FeII) are shown in green and the red line is the total fit.  The shaded yellow regions indicate the zone affected by strong sky line residuals  which are excluded from the fit.  \emph{Lower panel}: fit residuals, obtained as a difference between observed  and model spectra.}
 \label{fig:spectrum}
   \end{figure*}

Figure \ref{fig:spectrum} shows the H-band spectra extracted from the nuclear region  of each QSO from an aperture of 0.25\arcsec$\times$0.25\arcsec. The spectra clearly show  the broad \hb\ and the strong emission line doublet [O{\sc iii}]$\lambda\lambda5007,4959$. In addition to these components, weak FeII emission lines are also visible in two out of the six QSOs (LBQS0109 and HE0251). The  asymmetric \oiii\ profiles suggest  the presence of ionised outflows  in most of the targets (LBQS0109, 2QZJ0028, HB8905, HB8903 and HE0251). Indeed the presence of a  prominent blue wing is the typical signature of high velocity gas moving toward the line of sight.
The \hb\ emission line shows  asymmetric velocity profiles as well as, in some cases,  two distinct emission peaks (i.e HB8903). Indeed, in all spectra  the  \hb\ line is a sum of two different components: a very broad ($FWHM>2000$ km/s) one which is associated to the BLR emission and a fainter, narrower ( FWHM$\sim$ 500-1200 km/s) one likely associated to NLR emission. 

\subsection{Fitting procedure}\label{sec:fitting}
 In order to understand the dynamic and the main properties of the ionised outflows, we performed a kinematical analysis on the forbidden lines. 


At first, we extracted the spectrum of each QSO from a nuclear region of 0.25\arcsec$\times$0.25\arcsec\ where the signal-to-noise (S/N) is  highest. We fitted simultaneously the continuum, \hb\ and [O{\sc iii}] emission lines  by minimizing $\chi^2$  using the IDL routine MPFIT (\citealt{Markwardt:2009}). 

The \oiii\ line profile is very complex in these objects and a single Gaussian function is not able to reproduce the asymmetric velocity profiles, so we fitted  the emission-line doublet using either  multiple Gaussian components or a broken power-law convolved with a Gaussian distribution. Initially, we fitted the emission line using two Gaussian components, one broad (FWHM $> 1000$ km/s) and one narrow (FWHM $< $ 1000 km/s), and, when necessary, added a third component to minimise the $\chi^2$ value. Since in a few cases the \oiii\ emission line exhibits a very asymmetric line profile with  blue wings extended to $|v| >$1000 km/s, we replaced the multi-Gaussian components with a broken power-law profile convolved with a Gaussian distribution. We did not add  additional Gaussian components in order to avoid an unnecessary large  number of free fitting parameters. 
We do not attribute any physical meaning to the individual fitting component and we measure  gas kinematics by analysing the total line profile. Therefore, using either  multiple Gaussian components  or a broken-power law  does not change the results of this work, provided that the quality of the fit is similar. 
The two emission lines of the [O{\sc iii}] doublet, originating from the same upper level,  were  fitted imposing the same central velocity and velocity dispersion, with the intensity ratio $I(5007)/I(4959)$ fixed at $\sim$3. 

\hb\  is well described by two components, a broad (BLR) and a narrow one (NLR).  We used a broken power-law profile for the very broad component (FWHM > 2000 km/s) since it usually provides a good fit to QSO broad emission lines (\citealt{Nagao:2006}). The model used for the  \hb\ profile of 2QZJ0028 shows a broad red wing that is likely due to continuum or FeII emission, so we do not attribute any physical meaning to this red-shifted emission. A similar profile is visible in Figure 1 of \cite{Cano-Diaz:2012}.  Since the narrower \hb\ is weak and sometimes only marginally detected, it is not possible to reliably constrain its profile and kinematics and therefore  we assume that it has the same average velocity and velocity profile as \oiii. 

Finally we used a power-law  for the continuum and, for those spectra showing    FeII  emission, we used the FeII template from \cite{Tsuzuki:2006}.
These best-fitting solutions are used as first  guesses in the  pixel-to-pixel fitting of the kinematical analysis described in the following.

\begin{figure}
	\centering

	\includegraphics[width =.45\textwidth]{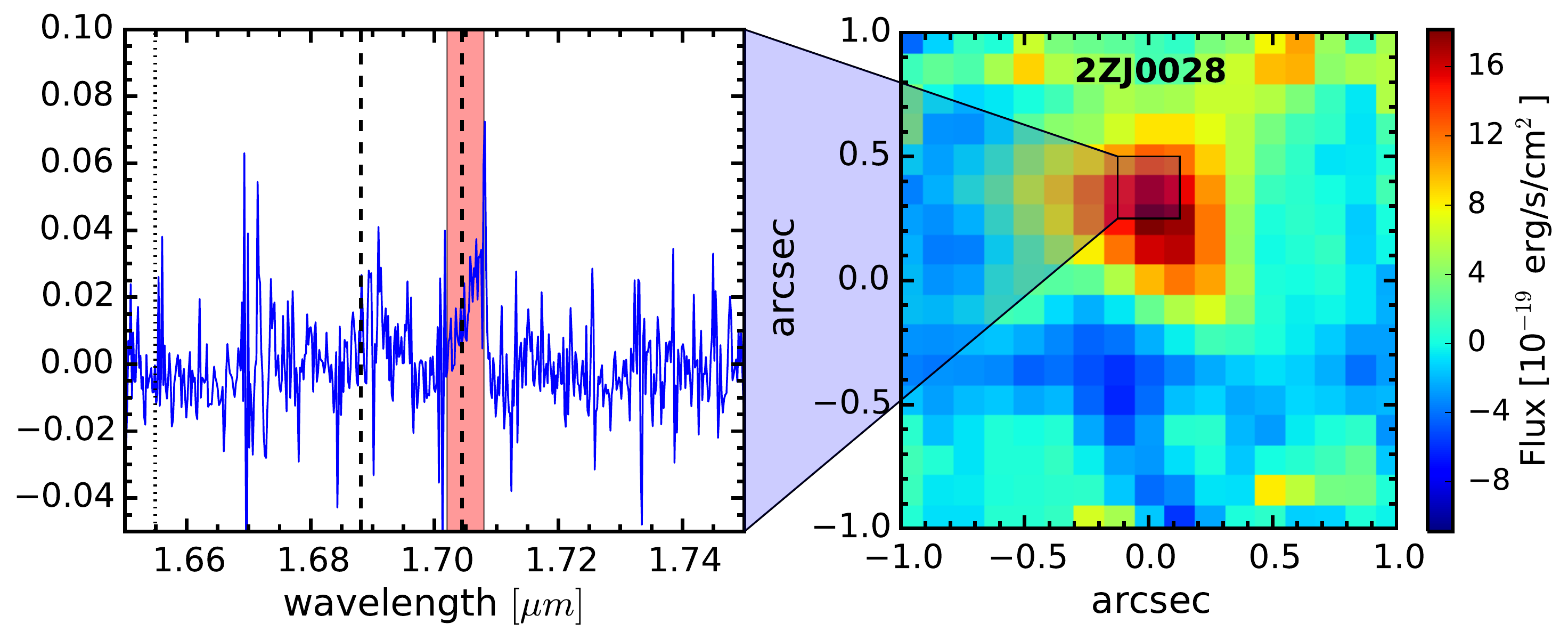}
	\includegraphics[width =.46\textwidth]{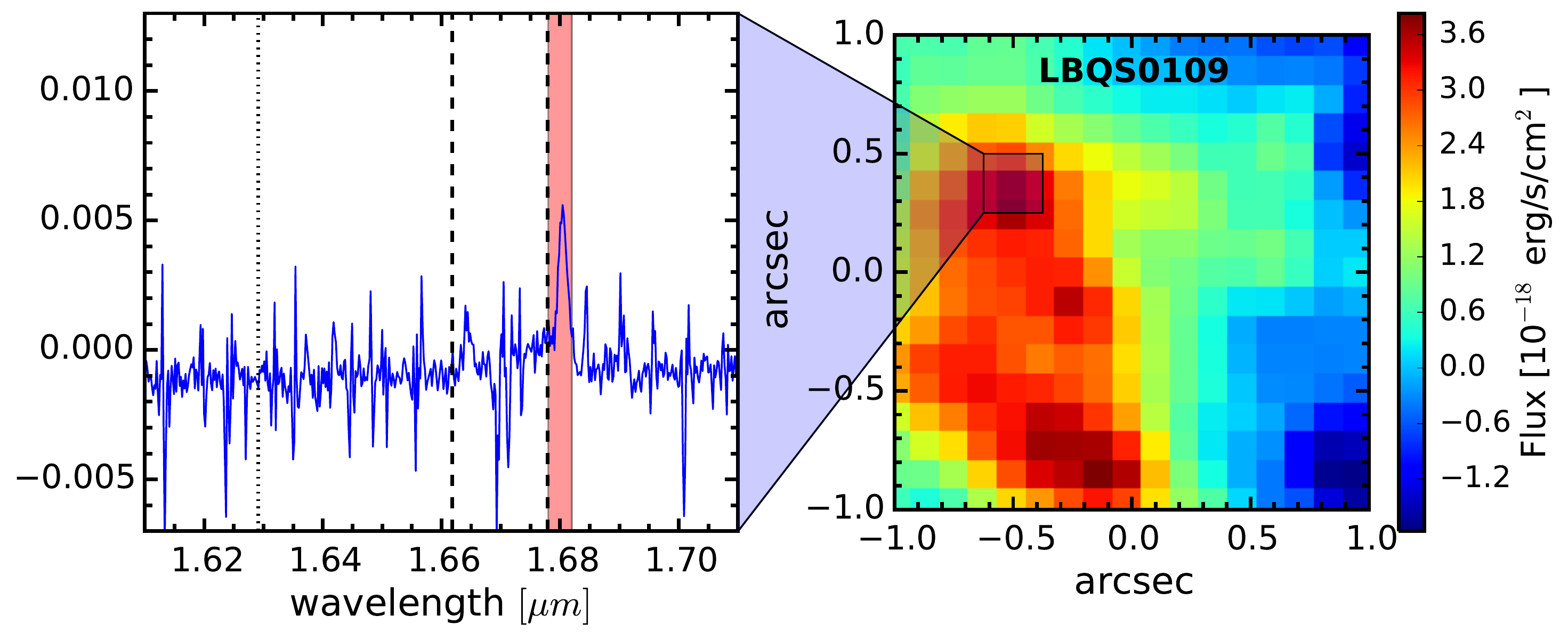}
	\includegraphics[width =.45\textwidth]{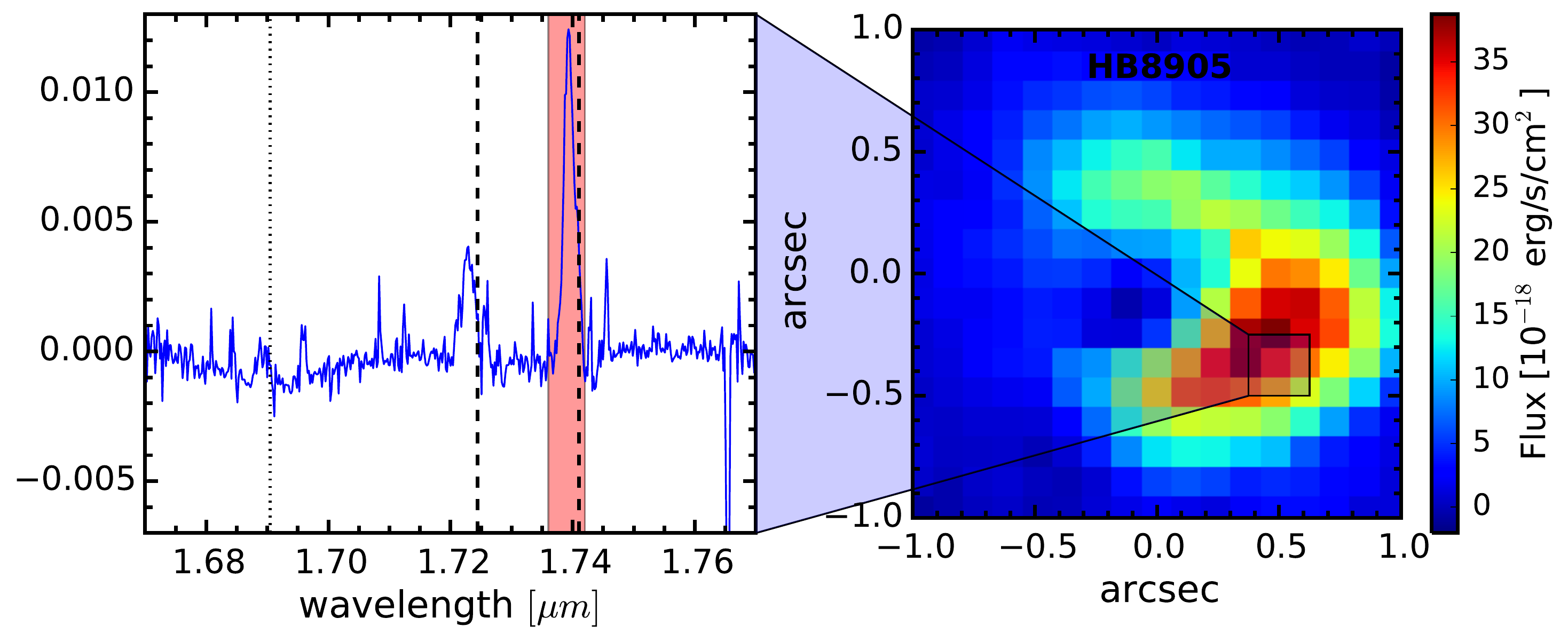}
	\includegraphics[width =.45\textwidth]{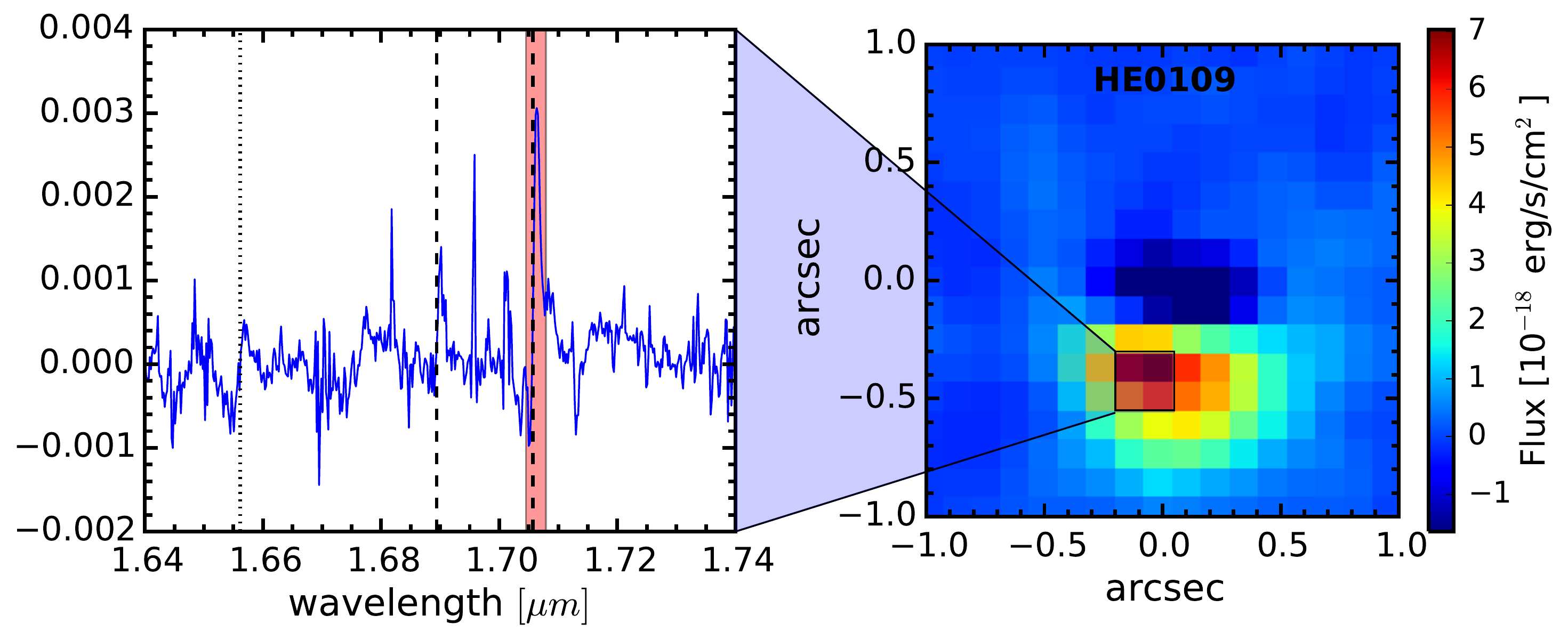}
	\includegraphics[width =.45\textwidth]{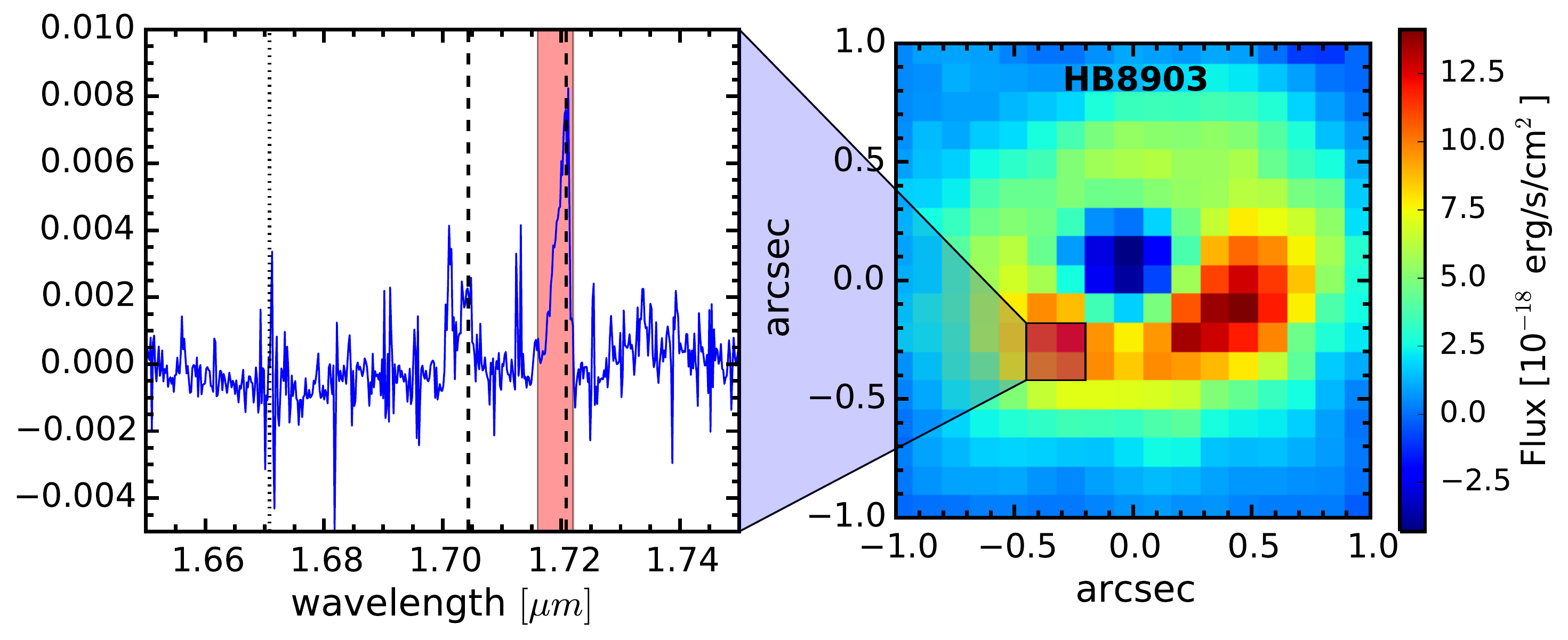}

	\includegraphics[width =.45\textwidth]{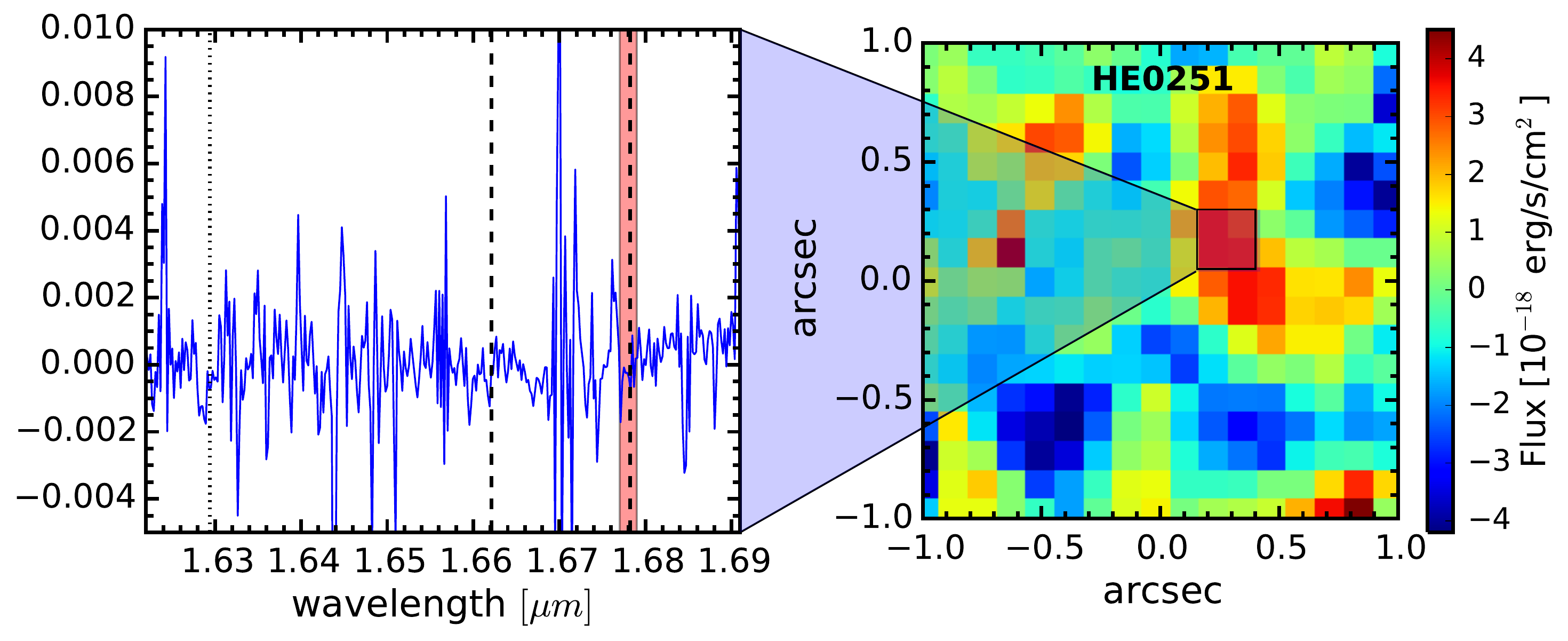}

    \caption{Residuals from the pixel-per-pixel fitting assuming that the QSO emission is not spatially resolved. Left panels:  residual spectra extracted from a region of 0.25\arc$\times$0.25\arc, where the residual map, obtained by collapsing the \oiii\ spectral channel, shows likely the presence of a spatially resolved emission. Dashed lines indicate the wavelength of the doublet \oiii\ and the dotted line shows the \hb\ position. The red  shaded region denotes the wavelength range over which the residual \oiii\ emission has been integrated to produce the maps shown on the right. Right panels:  residual maps obtained by collapsing the spectral channels corresponding to the residual \oiii\ emission, as shown by the red region in the left panels. In the first five maps the presence of a clear \oiii\ residual emission suggests that the emitting region is spatially resolved. The ``noise'' residual map of HE0251 indicates that the sources is not resolved.}
    		\label{fig:residual_map}
   	\end{figure}

Before performing a spatially resolved kinematical analysis, we tested whether the kinematics of the ionised gas is  spatially resolved.
Following the method described in \cite{Carniani:2013}, we  analysed the flux residual maps obtained with a pixel-by-pixel kinematical fitting with the components just described, after assuming that the targets are spatially unresolved. In this case, the H-band spectrum  is expected to be the same in  any spaxel apart for a different normalisation.
The residual maps (Figure \ref{fig:residual_map}) at the wavelength range of \oiii\ indicates that the forbidden line emission   is kinematically resolved in all but one
of the quasars, i.e HE0251.
In fact, if the emission was unresolved, we would expect to observe both a spectral profile and a residual map consistent with the noise, as in the case of HE0251.
The flat residual at the wavelength range of \hb\ indicates that the broad
component of \hb\ is spatially unresolved, consistently with a \hb\ origin in the BLR. 
 
\begin{figure*}
   \quad\quad\quad\quad\quad\quad FLUX MAP	\quad\quad\quad\quad\quad MEDIAN VELOCITY MAP \quad\quad\quad $v_{10}$ MAP \quad\quad\quad\quad VELOCITY DISPERSION MAP 	\\
    \centering

\includegraphics[width =.9\textwidth]{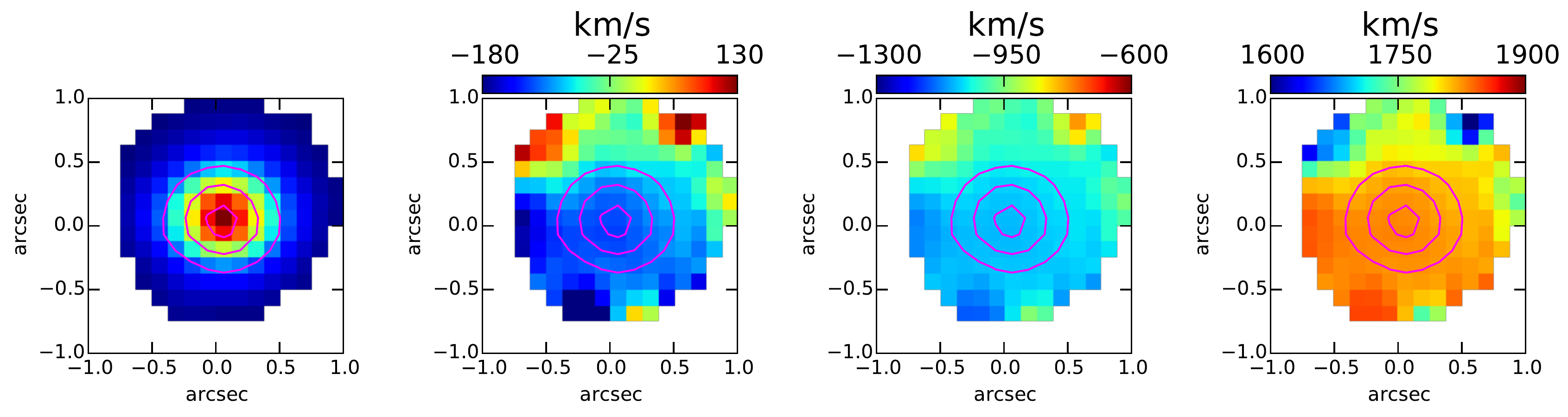}
\put(0,42){\rotatebox{90}{2QZJ0028}}  \\

\includegraphics[width =.9\textwidth]{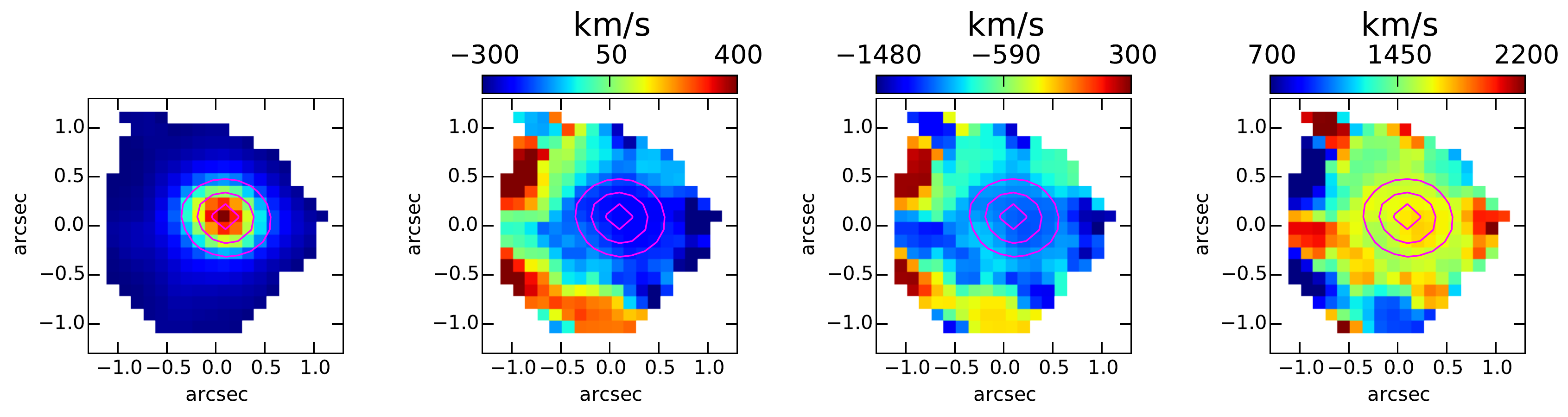} 
\put(0,42){\rotatebox{90}{LBQS0109}}  \\

\includegraphics[width =.9\textwidth]{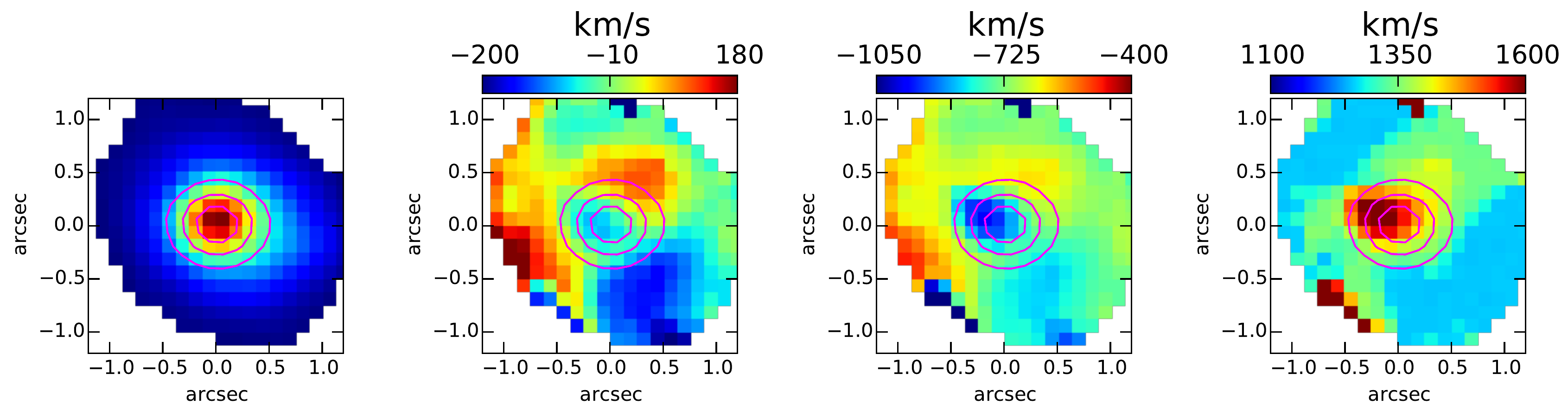} 
\put(0,42){\rotatebox{90}{HB8905}}  \\

\includegraphics[width =.9\textwidth]{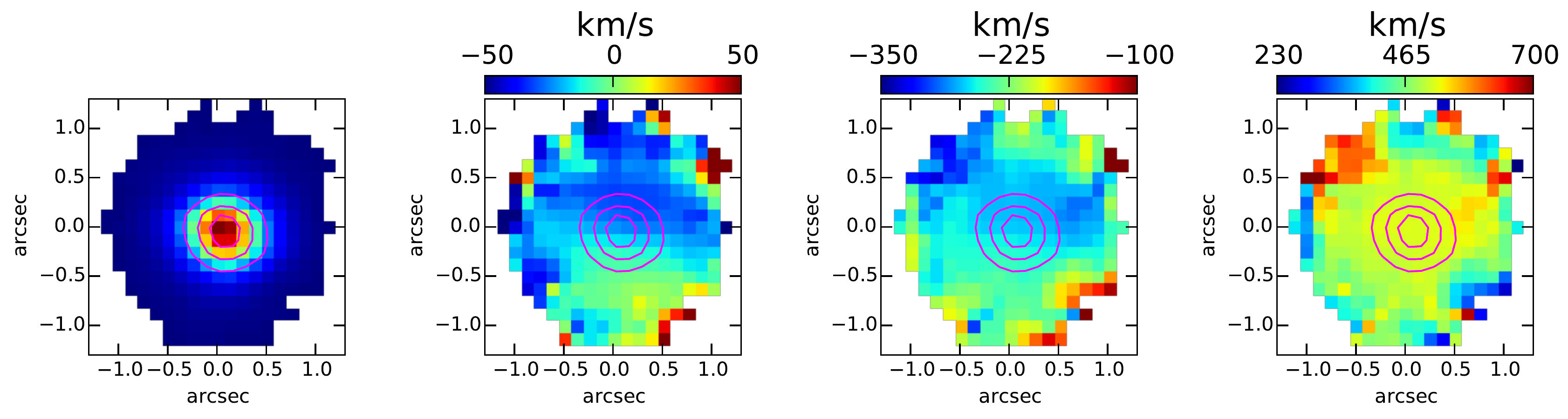} 
\put(0,42){\rotatebox{90}{HE0109}}  \\

\includegraphics[width =.9\textwidth]{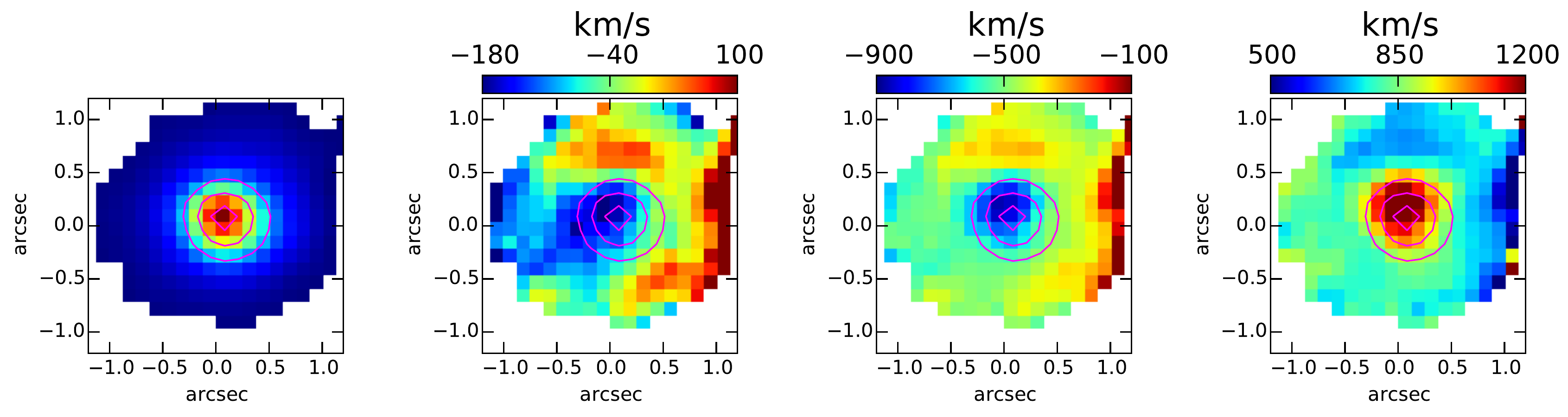} 
\put(0,42){\rotatebox{90}{HB8903}}  \\

\caption{\oiii\ flux, median velocity, $v_{10}$, and velocity dispersion map. The maps are obtained by selecting pixels with a SNR>2. The velocity maps are characterised by blue shifted regions with a large velocity dispersion. Contours represent the total \hb\ line surface brightness at 90\%, 50\%, and 30\% of the peak value}
\label{fig:velocity_maps}
	\end{figure*}

After testing that the  \oiii\ line emission is kinematically
resolved in five of the six sampled QSOs, we performed a pixel-by-pixel 
fitting on the data cubes allowing the spectral components to vary except for the broad \hb\ which is not spatially resolved.
 Figure \ref{fig:velocity_maps} shows the kinematical properties of the ionised gas as obtained from the  fit of the \oiii\ line:
\begin{itemize}
	\item zeroth moment map (i.e., line flux map; first panel);
	\item first moment map (i.e., median velocity map; second panel);
	\item blueshifted velocity map, $v_{10}$,  the velocity at the 10th percentile of the overall emission-line profile (third panel);
		fitted in each spatial pixel;
	\item line width  map , $W_{80}$, the velocity width of the line that contains 80\% of the emission-line flux (fourth panel); this is defined as $W_{80} = v_{90}- v_{10}$, where $v_{10}$ and $v_{90}$ are the velocities at 10th and 90th percentiles, respectively. For a Gaussian profile, $W_{80}$ is approximately the full width at half maximum (FWHM).
\end{itemize}
\noindent The observed kinematical maps are the results of the convolution of the intrinsic ones with the PSF of the observations.
In those cases where the \oiii\ emission line has been fitted with multiple Gaussians,  line profiles are based on the sum of all Gaussian components.
The maps
were obtained by selecting only those spatial pixels
with a S/N  equal to or
higher than 2. We defined the S/N as the ratio between 
the peak of the \oiii\ line and the rms of
the residuals.  Zero-velocities correspond to the peak position either of the 
 narrow (FWHM $<$ 1000 km/s) and strongest  Gaussian component or of the broken-power law one of the
\oiii\ profile estimated in the preliminary spectral
fitting. The zero-velocity wavelength was also used to refine the
redshift of each QSO (Table \ref{tab:summary}) providing the 
velocity of the host galaxy. The inferred redshift and \oiii\ line width are consistent, 
within the errors, with those estimated by \cite{Shemmer:2004} and
 \cite{Marziani:2009}.

The $v_{10}$ maps (Figure \ref{fig:velocity_maps}, third panel)
show strongly blue shifted regions, spatially associated with high velocity dispersion ($>$ 400 km/s, Figure  \ref{fig:velocity_maps}, forth panels).
The broad \oiii\ profile cannot be explained by a rotating gas component, which, in local star forming galaxies has typical
FHWM values of about $\sim$250 km/s. Moreover, the morpholgy  of the velocity maps , suggesting the presence of a conical  blue-shifted region, is completely different  from the typical ``spider'' diagram of a disc. Consequently both the \oiii\ profile and the velocity maps suggest that
in at least five out of six QSOs we detect ionised outflowing  gas with velocities $>300$ km/s.


We can detect only the blue-side outflows because the
red-side is likely to be obscured by dust in the host galaxy along the
line-of-sight. For this reason, the \oiii\ line emission is asymmetric with a prominent blue shifted wing.  While this is what commonly happens,  in some cases the particular
orientation of the line-of-sight with  respect to the source can result in redshifted outflows (e.g. \citealt{Rodriguez-Zaurin:2013,Bae:2014, Perna:2015}).

The next step is to understand the physical mechanisms driving away the ionised gas
out to a distance of a few kpc  from the centre of the host galaxy.

\section{Outflow properties}\label{sec:outflow}
In the following, we estimate mass, average velocity and radius of the high-velocity winds. We then infer mass outflow rate, momentum rate and kinetic power, and compare them with previous works
(\citealt{Greene:2012}, \citealt{Cicone:2014}, \citealt{Harrison:2014}, \citealt{Sun:2014}, \citealt{Brusa:2015}, \citealt{Cresci:2015} and \citealt{Feruglio:2015}).

\subsection{Outflow and radius}\label{sec:spectra}
Given the uncertainties on the driving mechanism, we used a simple model to estimate the physical properties of the outflow. In this model, the outflow is represented by a shell-like cloud ejected from the nucleus within a cone  and with a  filling factor equal to 100\%. 
We assume that the physical size of the outflowing material is smaller than the spatial  resolution typical of our datasets.  
This  model is defined  by the mass of the cloud, $M_o$, the distance between
 the cloud and the location of the AGN, $R_o$, and the averaged cloud velocity, $v_o$.
 The outflow mass rate is given by
\begin{equation}
\dot M_{o} = \frac{M_{o}}{\tau_{dyn}} = \frac{M_{o} v_{o}}{R_{o}}
\label{eq:rate}
\end{equation}
where $\tau_{dyn}$ is the dynamical time, i.e. the time taken by the ionised 
gas to reach a distance $R_o$  with an average velocity  $v_{o}$. We note 
that assuming either a shell or a uniformly filled cone with a filling factor equal to 1 (e.g \citealt{Maiolino:2012})  
changes the $\dot M_{o}$ estimate by a factor of 3. Since this factor is constant, it does not affect the main conclusions of this work but we will take  it into account when comparing our results with previous ones.

 The extension of an outflow ($R_{o}$) is usually trivially estimated from the observed flux or velocity maps (e.g. \citealt{Harrison:2014}, \citealt{Cresci:2015}). However, in our cases,
the \oiii\ emission  is only marginally spatially resolved in each QSO,  therefore the  kinematical maps in Figure \ref{fig:velocity_maps} are affected by PSF smearing and the sizes of
the regions showing blue-shifted emission (Fig. \ref{fig:velocity_maps}) do not directly  provide the parameter $R_o$ to be used in equation \ref{eq:rate}.
Furthermore, the sizes of the outflowing regions  are affected by the choice of S/N threshold adopted in plotting the maps. In particular,
assuming a S/N threshold larger than 2, the blue regions would be
smaller than those shown in Figure~\ref{fig:velocity_maps}.  For these reasons, we cannot estimate $R_{o}$ from our observations  by using the flux or velocity maps as in previous works,  where  emission lines are clearly spatially resolved. Therefore, we decided to estimate
 $R_o$ by using spectroastrometry, 
 which allows position measurements on scales smaller than the spatial resolution of the observations. Spectroastrometry consists in measuring the photo-centroid  in each velocity channel. 
If outflowing ionised gas is moving away at a distance $R$ from the QSO, we expect that
the centroid of light emission extracted from  blue-shifted velocities channels  of \oiii\ will be displaced of the same amount $R$ with respect to the QSO  position, identified by the BLR and/or continuum emission.
In  Section \ref{sec:simulation}, by means of a simple simulation, we show that spectroastrometry can, indeed, provide information on the outflow position at scales that are significantly smaller  than the limit imposed by the spatial resolution of the observations.

We applied the spectroastrometry technique to the \oiii\  line emission after subtracting the best-fit model of the continuum, broad \hb\ and FeII emission.  To maximise  the signal-to-noise ratio for our measurements and minimize the uncertainty due the spectral resolution, we rebinned the spectra by 3 velocity channels ($\sim 105$ km/s). The centroid of the \oiii\ emission in each rebinned velocity channels was, then, estimated by a 2-dimensional Gaussian fitting. The QSO position was estimated by applying the same spectroastrometry technique to the the continuum and the broad \hb\ emission. 

\begin{figure}[tbh]
   \centering
\includegraphics[width =0.445\textwidth]{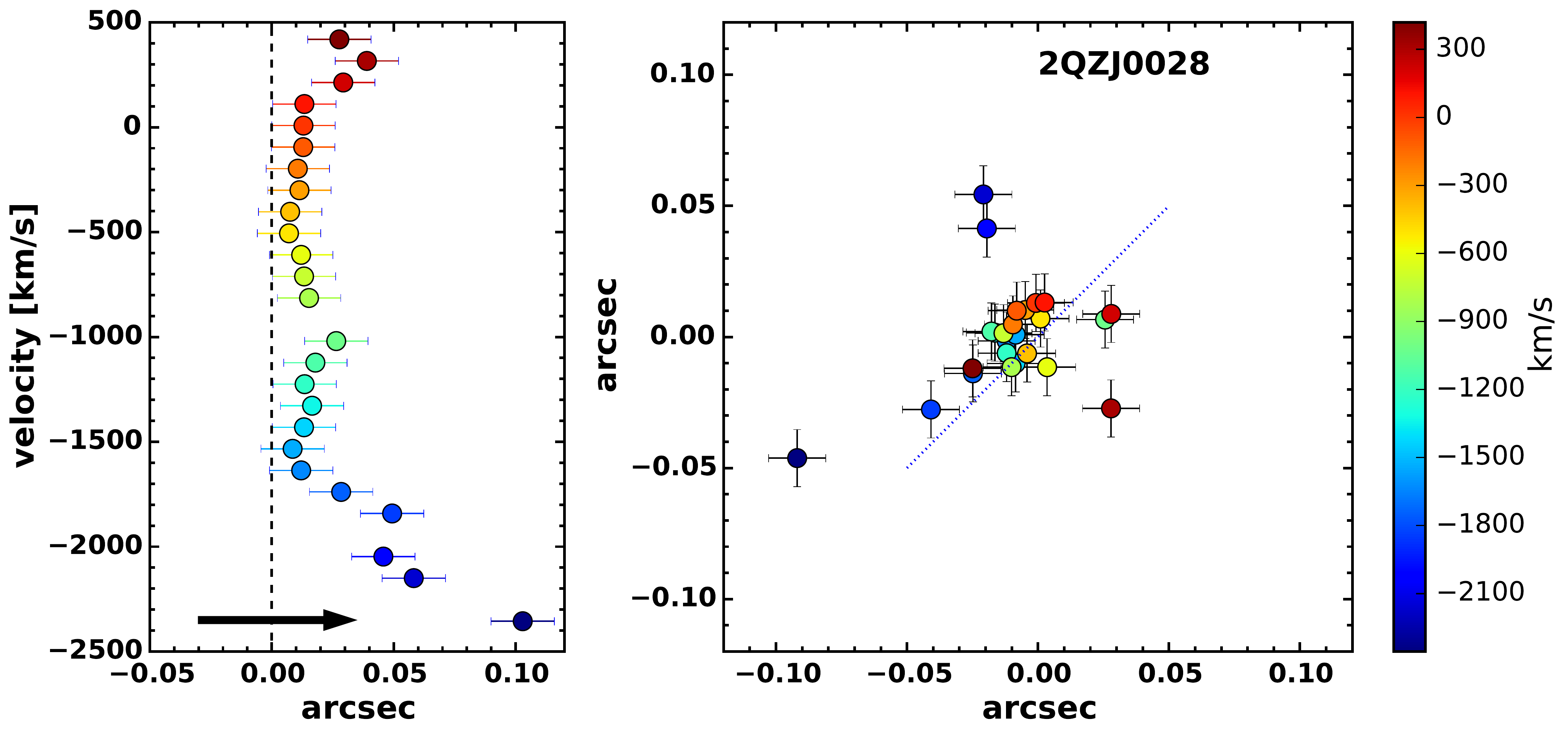}
 
\includegraphics[width =0.445\textwidth]{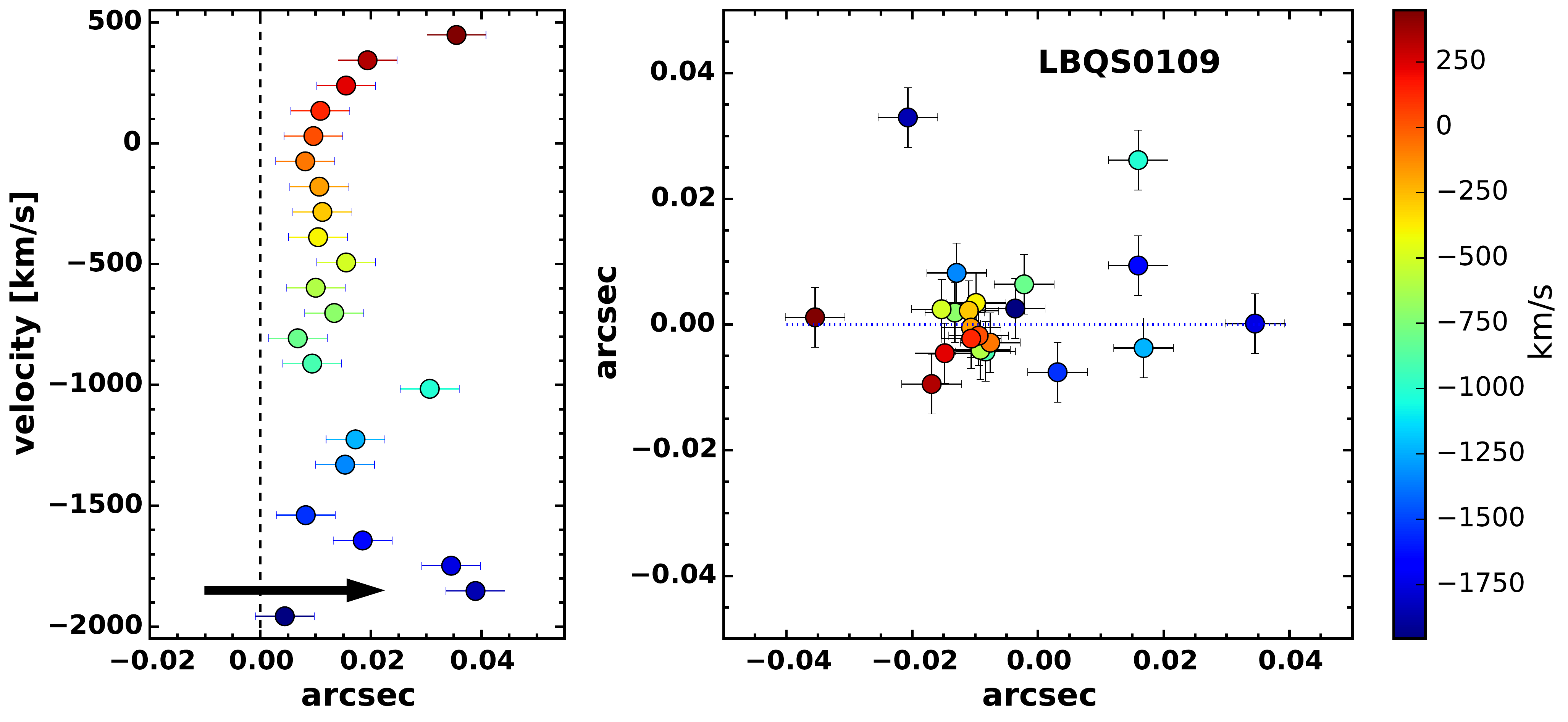} 

\includegraphics[width =0.445\textwidth]{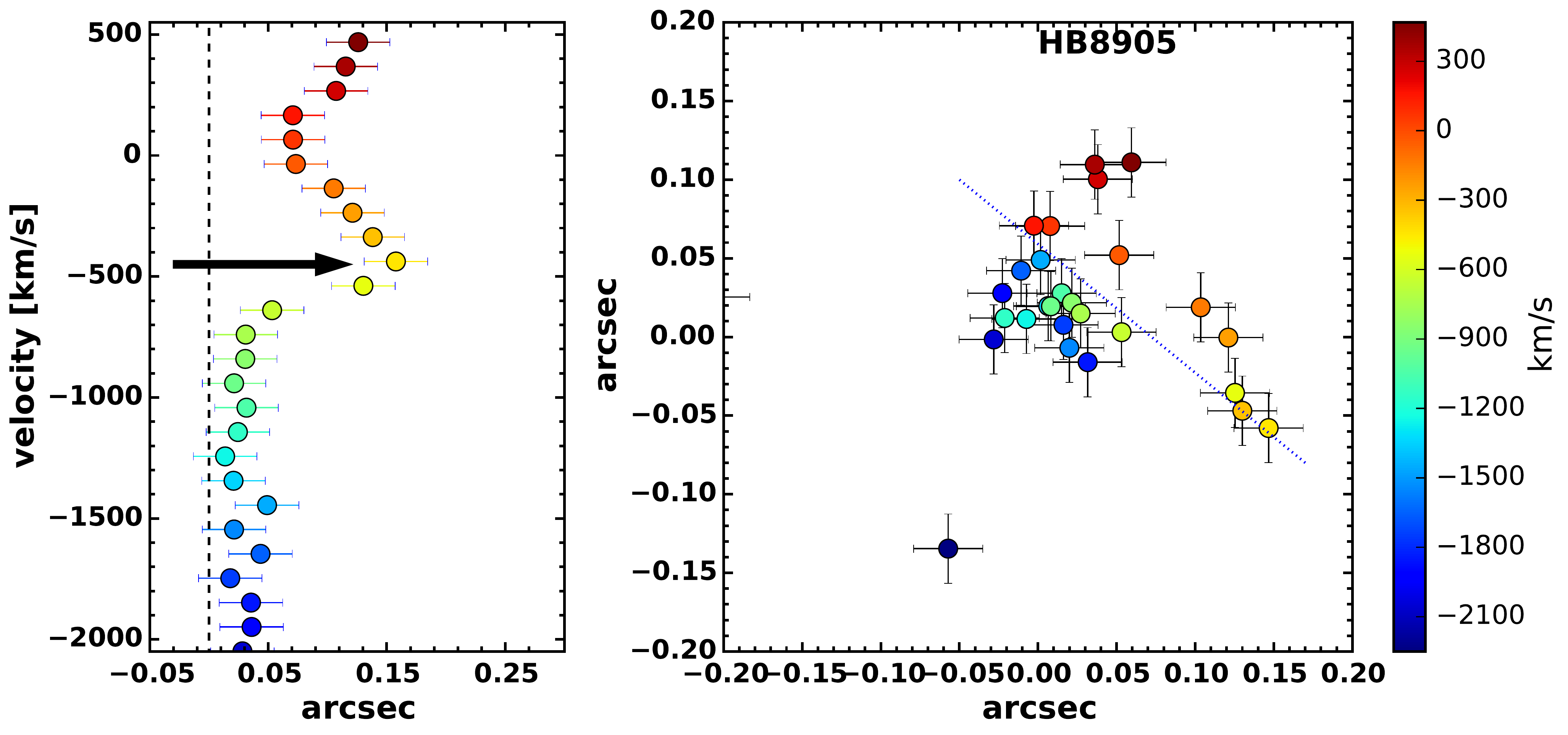}

\includegraphics[width =0.445\textwidth]{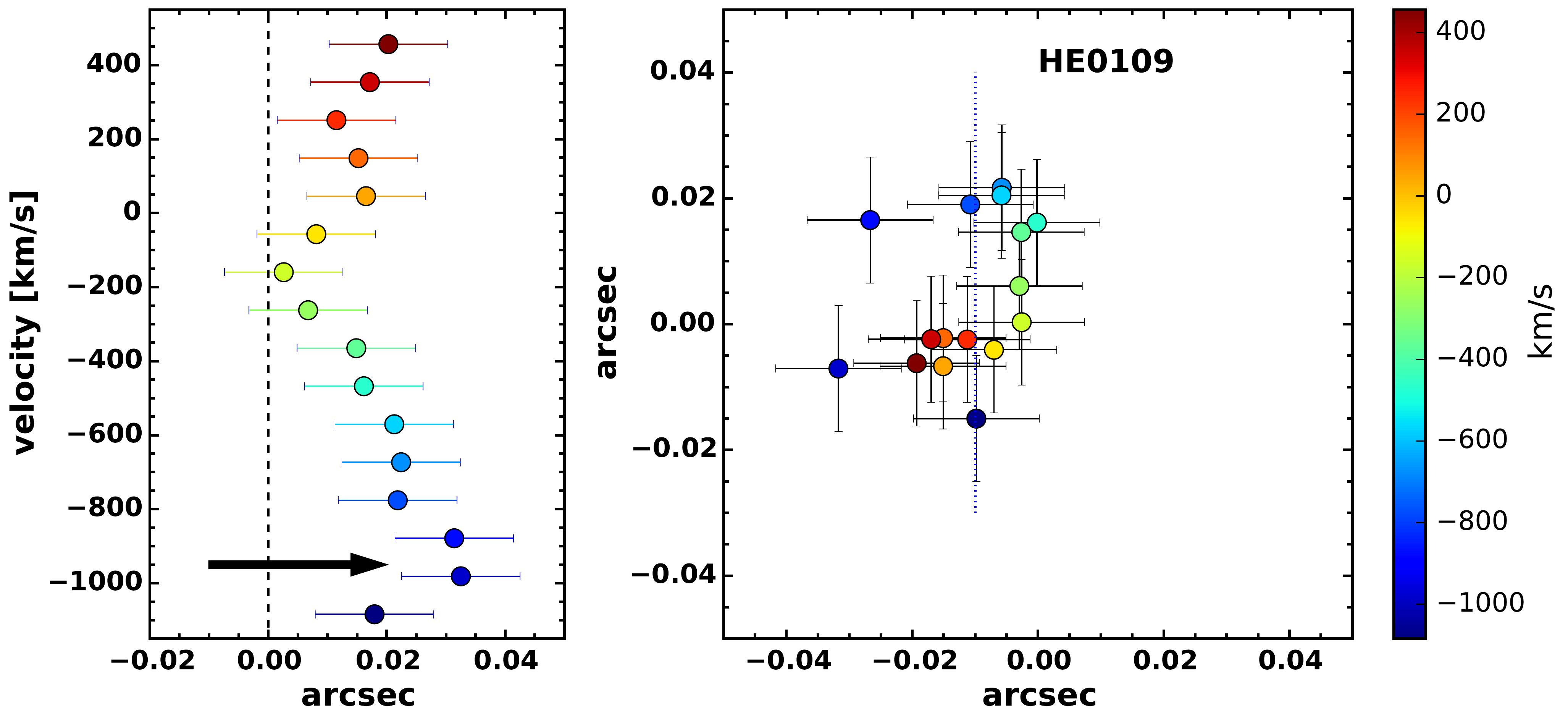} 

\includegraphics[width =0.445\textwidth]{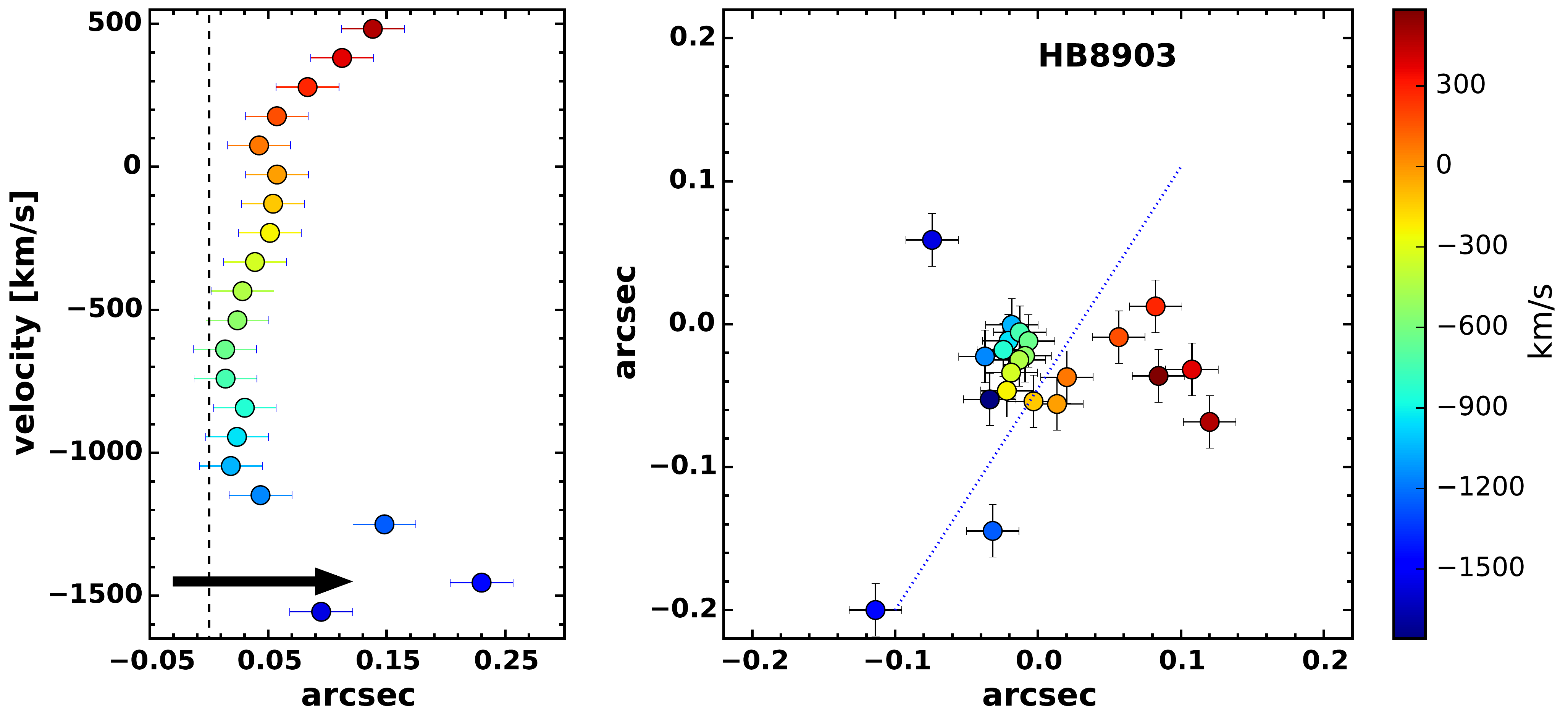} 

 \caption{ Left panels. The points show the \oiii\ velocity $v$ versus the distance $R$ of the \oiii\ photocenter from the continuum (indicated by the dashed line). 
The arrows indicate the  velocity $v_o = v(R_o)$ corresponding to the distance $R_0$. Right panels. The \oiii\ photocenter position in the field of view. Symbols are coloured according to their velocity (the velocity scale is 
reported in the colour bar). The dotted line indicates the project direction of the outflow as inferred by comparing the spectroastrometry results with velocity maps (Fig. \ref{fig:velocity_maps}).
 }
\label{fig:spectroastrometry}
   \end{figure}

The results of the spectroastrometric analysis are displayed in Figure \ref{fig:spectroastrometry}. In the left panels, we plot the distance $R$ of the emission line photocenter from the continuum one as a function of velocity $v$. The right panels show the photocenter position on the sky in each velocity channel.
The reliable spectroastrometric measurements for the \oiii\ emission line were selected to satisfy the following criteria: 
\begin{itemize}
\item signal-to-noise larger than 1.5 for the line flux in each spectral channel of the rebinned spectra extracted from a $0.25\arcsec\times0.25\arcsec$ 
\item FWHM of the 2-dimensional Gaussian equal or larger than that of the PSF of the observations
\end{itemize}

In all targets, the \oiii\ blue wings centroids are displaced  at least 0.05\arcsec, i.e. 0.4  kpc, from the continuum position and, in some cases, we observe  an  offset  in the red wavelengths as well. This latter offset may be caused by three reasons: (i) the continuum is not well subtracted during the kinematic analysis; (ii) the \oiii\ emission line associated to outflows is so large that the displacement  is slightly observable also in the red wings; (iii) the red wings are emitted by the receding outflows not completely obscured by dust in the disc. An additional explanation, which will be discussed in a companion paper is that in two out of six QSOs a fraction of the red \oiii\ line emission is associated with star formation in the host galaxy causing a  displacement from the continuum centre. 
 
Since the position of the photocenter  at high blue velocities reveals the presence of the extended  ionised outflow, we adopt $R_o$ equal to the largest distance measured for the approaching gas, and $v_o = v (R_0)$ as measured from our maps (see arrows in Figure~\ref{fig:spectroastrometry}).
Since the spectra have been rebinned by 3 velocity channels, the error on $v_0$ due to the spectral resolution typical of our datasets is negligible.   
$R_o$ and $v_o$ for each quasar are listed in Table 2.


\begin{table*}
\caption{Outflow properties}           
\label{tab:outflow}      
\centering          
\begin{tabular}{l c c c c c c  c  c }    
			 \\
QSO &   $v_o$	& $R_o$	& ${\rm Log(L_{[OIII]}^{outflow})}$	& ${\rm Log(L_{H\beta}^{outflow})}$ &
 $M^{outflow}_{\rm [OIII]}$ &  $M^{outflow}_{\rm H\beta}$	& $  \dot M_{\rm [OIII]}$ &  $\dot M_{\rm H\beta}$ \\ 

	& [km/s]	& [kpc]	& 							& 								&
   [$10^7$ \msun]    &  [$10^7$ \msun]  & [\sfr ] & [\sfr ]  	\\
\hline
LBQS0109  & 1850  & 0.4  & 43.17 &  42.10  & 1.2 & 2.2 & 60 & 110 	\\
2QZJ0028& 2300  & 0.7  & 43.68 & 43.07 & 3.8 & 20 & 140 & 700    \\
HB8905  & 500   & 1.3 & 43.95 & 42.99 & 7.1 & 16.8 & 30 & 75    \\
HE0109    & 900   & 0.4 & 43.75 &  42.71 & 4.5 & 8.6 & 110 & 210   \\
HB8903 & 1450  & 1.9 & 42.95 & 41.8 & 0.7 & 1.2 & 6 & 10 \\
HE0251   &   -   &  -  & 43.82 & 42.55 & 5.3 & 6.1 & - & - \\
\hline
\multicolumn{9}{l}{\tiny \emph{Notes}:  The outflow masses are estimated assuming a ${\rm T_e \sim 10^4}$ K and a ${\rm n_e \sim 500 cm^{-3}}$.} 
\\
\end{tabular}	
\end{table*}

%

\subsection{Outflow Mass }\label{sec:mass}

A big challenge in estimating the mass of the wind is caused by its multiphase nature. Indeed, only a fraction of the mass of the outflows is in the warm ionised phase traced by \oiii. Recent works estimate the mass of ionised outflows either using \hb\ emission line (e.g \citealt{Liu:2013}; \citealt{Harrison:2014}) or using \oiii\ line (e.g  \citealt{Cano-Diaz:2012}). So far, it is not clear which is the best tracer of ionised gas powered by AGN feedback. To  compare two different measurements carried out with these two different tracers,  we discuss how the mass of ionised outflows can be constrained through the observations of \oiii\ and \hb.

The \oiii\ luminosity  is given by
\begin{equation}
L_{\rm [OIII]} = \int_{V} fn_en(O^{2+})j_{\rm [OIII]}(n_e,T_e)dV
\label{eq:loiii}
\end{equation}
where $f$ is the filling factor,  $n_e$ the electron density, $n(O^{2+})$ the density of $O^{2+}$ ions and $j_{[OIII]}(n_e,T_e)$ the line emissivity. $n(O^{2+})$ can be written as
$$
n(O^{2+}) = \left[\frac{n(O^{2+})}{n(O)}\right]\left[\frac{n(O)}{n(H)}\right]\left[\frac{n(H)}{n_e}\right]n_e
$$
and,  with a reasonable assumptions, ${n(O^{2+})}={n(O)}$, 
$$
n(O^{2+}) \simeq (6.04\times10^{-4}10^{[O/H]-[O/H]_{\odot}})\times(1.2)^{-1}\times n_e
$$
where  $[O/H]-[O/H]_{\odot}$ is the metallicity relative to solar  with  a solar oxygen abundance of $[O/H]_{\odot} \sim8.86$ \citep{Centeno:2008}.
The factor $(1.2)^{-1}$ takes into account a 10$\%$ number density of He atoms with respect to H atoms:
$$
n_e \approx n(H) + 2n(H_e) = n(H)+2\times0.1\times n(H) = 1.2n(H)
$$
Assuming a typical temperature ($T_e\simeq 10^4 K$) and electron density ($n_e\simeq500$   cm$^{-3}$) for the NLR the line emissivity is
$$
j_{\rm [OIII]}= 3.4\times10^{-21} {\rm  \ erg  \ s^{-1} cm^{-3}}
$$
This line emissivity was estimated making use of PyNeb \citep{Luridiana:2015}.  
Therefore, equation (\ref{eq:loiii}) can be rewritten as:

\begin{equation}
L_{\rm [OIII]}=  6.0\times10^{-4}f10^{[O/H]-[O/H]_{\odot}}j_{\rm [OIII]}<n_e^2> V
\label{eq:loiii2}
\end{equation}
where $<n_e^2>$ is the volume-averaged squared density.
The gas mass can be expressed as

\begin{equation}
 M \simeq \int_V f \overline{m} n(H) dV \simeq f m_p <n_e> V
 \label{eq:mass}
\end{equation}
 where $\overline{m}$ is the average molecular weight, $m_p$ is the proton mass, and  we have taken into count that  a 10\% number density of He atoms with respect to H atoms:
$$
\overline{m}n(H) \approx \frac{m_p n(H)+4m_p n(He)}{ n(H)+n(He)}  \frac{n(H)}{n_e}n_e 
$$
$$
\approx \frac{n(H)+0.4n(H)}{ n(H)+0.1n(H)} m_p (1.2)^{-1}n_e  \approx 1.2 m_p (1.2)^{-1}n_e = m_p n_e
$$
Finally, combining  equation (\ref{eq:loiii2})   and  equation (\ref{eq:mass}) we  get:
\begin{equation} 
M_{\rm [OIII]}= 1.7\times10^3 \frac{ m_p C L_{[OIII]}}{10^{[O/H]-[O/H]_{\odot}}j_{[OIII]}<n_e>}
\label{eq:mass_oiii}
\end{equation}
where $C = <n_e>^2/<n_e^2>$. For $T_e = 10^4 $ K and $n_e = 500 \ {\rm cm^{-3}} $ we obtain a mass of:
$$
M_{\rm [OIII]}= 0.8\times10^8\msun \left(\frac{C}{10^{[O/H]-[O/H]_{\odot}}}\right) \left(\frac{L_{[OIII]}}{10^{44} erg/s}\right) \left(\frac{<n_e>}{500 cm^{-3}}\right)^{-1}
$$
Note that the gas mass is sensitive to temperature and density of the clouds.

Now we similarly derive $M$ from \hb. The \hb\ luminosity can be expressed as

\begin{equation}
L_{H\beta} = \int_V f n_e n_p j_{H\beta}(n_e,T_e) dV \simeq 1.2^{-1} f j_{H\beta}(n_e,T_e) <n_e^2>V
\label{eq:lhb}
\end{equation}
where $j_{H\beta}$ is line emissivity and $n_p$ is the proton density that can be written as 

$$
n_p = \left[\frac{n(H)}{n_e}\right]n_e = (1.2)^{-1}n_e
$$
At the typical temperature and density of the NLR, the line emissivity of \hb\  also estimated with PyNeb is 
$$
j_{H\beta} = 1.2\times10^{-25} {\rm  \ erg  \ s^{-1} cm^{-3}}
$$
Combining equation (\ref{eq:mass})  with equation (\ref{eq:lhb}) of $L_{H\beta}$ we get
\begin{equation}
M_{H\beta} \simeq 0.8 \frac{m_p C L_{H\beta}} {j_{H\beta} <n_e>}
\label{eq:masshb}
\end{equation}
For $T_e \sim 10^4 $ K and $n_e \sim 500 \ {\rm cm^{-3}} $ we obtain a mass of:
$$
M_{H\beta} = 1.7\times10^{9}\msun \ C \left(\frac{L_{H\beta}}{10^{44} erg/s}\right) \left(\frac{<n_e>}{500cm^{-3}}\right)^{-1}
$$
Comparing the two mass at the same temperature and density, we get 

\begin{equation}
\frac{M_{[OIII]}}{M_{H\beta}} \simeq 0.05 \frac{L_{[OIII]}}{L_{H\beta}}
\label{eq:ratio}
\end{equation}
    with [O/H] = ${\rm [O/H]_{\odot}}$, T$_e \sim 10^4$ K and n$_e \sim$ 500 cm$^{-3}$, the same values which will be assumed in the following analysis.
 Usually the \oiii/\hb\ ratio measured in Seyfert galaxies is of the order of $\sim10$ providing  $M_{[OIII]}/M_{H\beta}\sim 0.5$. This apparent discrepancy  is the consequence of the  different volumes 
from which \oiii\ and \hb\ are emitted: in the above example the volume of \hb\ emitting gas is $\sim2$ larger than that of \oiii. 
Therefore, \hb\ emission traces a larger amount of  ionised gas mass  than \oiii. 
The mass estimated from \oiii\ will then be considered as a lower limit on the ionised gas mass. 

\subsection{Simulations of spectroastrometric observations of outflows}\label{sec:simulation}
In  Section \ref{sec:spectra} we have described our method to estimate $R_o$, 
which  is not affected by the spatial resolution of the data and on the S/N threshold used to map velocities. 


 In the following we  present   results  of simple simulations of outflowing gas, validating  the spectroastrometric method. Note that the purpose of this work is not to build an outflow model to fit our observations, but the only purpose of this simulation is to show that the extension of the flux or velocity maps when the source is marginally resolved  can provide wrong estimates of the outflow extension due to PSF smearing. In fact
we will also show that an ionised clump with a given velocity and located a few kpc away from the AGN can appear as a blue-shifted region extended over $\sim$ 5-10 kpc  in velocity maps, which is  larger than the real outflow extension. 

 
 \begin{figure}
   \centering
\includegraphics[width =0.3\textwidth]{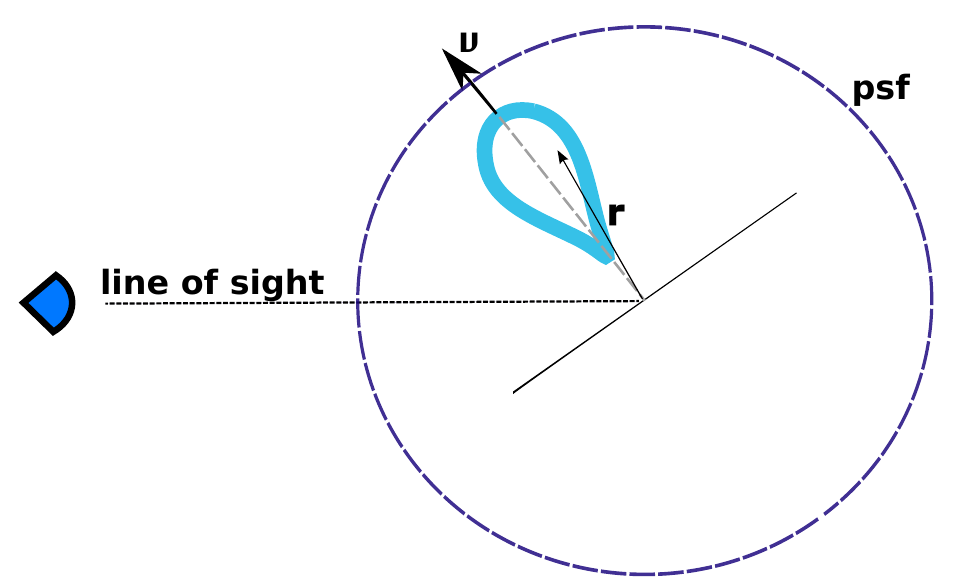}
 
\caption{Cartoon showing the basic structure of our model. The outflow (solid blue curve) is perpendicular to the galaxy plane. $\vec{v}$  indicates the direction of the outflow and $\vec{r}$ is the distance from the QSO. This model assumes that the outflow is marginally resolved but it is not larger than the PSF dimension (dashed purple line). }
\label{fig:toy_model}
   \end{figure}

We considered a simple model of an outflow (Fig. \ref{fig:toy_model}) based on the following assumptions.
\begin{enumerate}
\item  The outflow has a biconical geometry with the axis approximately more or less perpendicular to the host galaxy disk; indeed,  the AGN wind, although  roughly isotropic, cannot propagate through the galaxy disk because of its high density.
\item   The receding half cone is not observed because its emission is absorbed by the dust  in the host galaxy. This geometry best explains the asymmetric \oiii\ profile with a prominent blue wing over 1000~km/s.
\item The surface brightness distribution of the outflow is parameterised by 
$$I(\vec{r}) = I_0\rm{e}^{-\vec{r}/R_o^{model}}$$ 
where the vector $\vec{r}$ is the distance from the QSO.
Note that the surface brightness distribution and the opening angle of the cone are not fundamental for these simulations, since the  spatial resolution of our data is not high enough to resolve the surface brightness profile. We assume an opening angle of  30 degrees. $R_o^{model}$ identifies the photo-center of the surface brightness of the ionised cloud. $I_0$ is proportional to the  outflow mass $M_o^{model}$ defined in equation (4).
 \item The ionised gas in the outflow has an average velocity $v_o^{model}$ and a velocity dispersion $\sigma_o^{model}$.
 \end{enumerate}
Therefore the input parameters are the outflow mass $M_o^{model}$,  the distance R$_o^{model}$ of the ionised clouds from the centre of the  QSO,  the average velocity $v_o^{model}$ and  the velocity dispersion $\sigma_o^{model}$ of the ionised gas. 

In order to reproduce the \oiii\ line profile of LBQS0109 we simulated a QSO with an   outflow characterised by  $M_o^{model}$ =  $1.0\times10^{7}$ \msun\ (eq. 5), R$_o^{model}$ = 0.05 arcsec (i.e $\sim$0.4 kpc), $v_o^{model}$ = 1500 km/s  and $\sigma_o^{model}$ = 500 km/s. We also considered a point source at the location of the AGN describing the gas at the systemic velocity with velocity dispersion $\sim 400$ km/s and an amplitude scaled as to reproduce the observed line profile. 
The total \oiii\  emission was then projected onto the sky plane, assuming a galaxy disc inclination of  10 degrees, i.e. almost face on as  typical of QSO. Finally we convolved the surface brightness maps in each spectral channel with the PSF and added  Gaussian noise to match  the  sensitivity of our observations.

 \begin{figure}
   \centering
\includegraphics[width =0.4\textwidth]{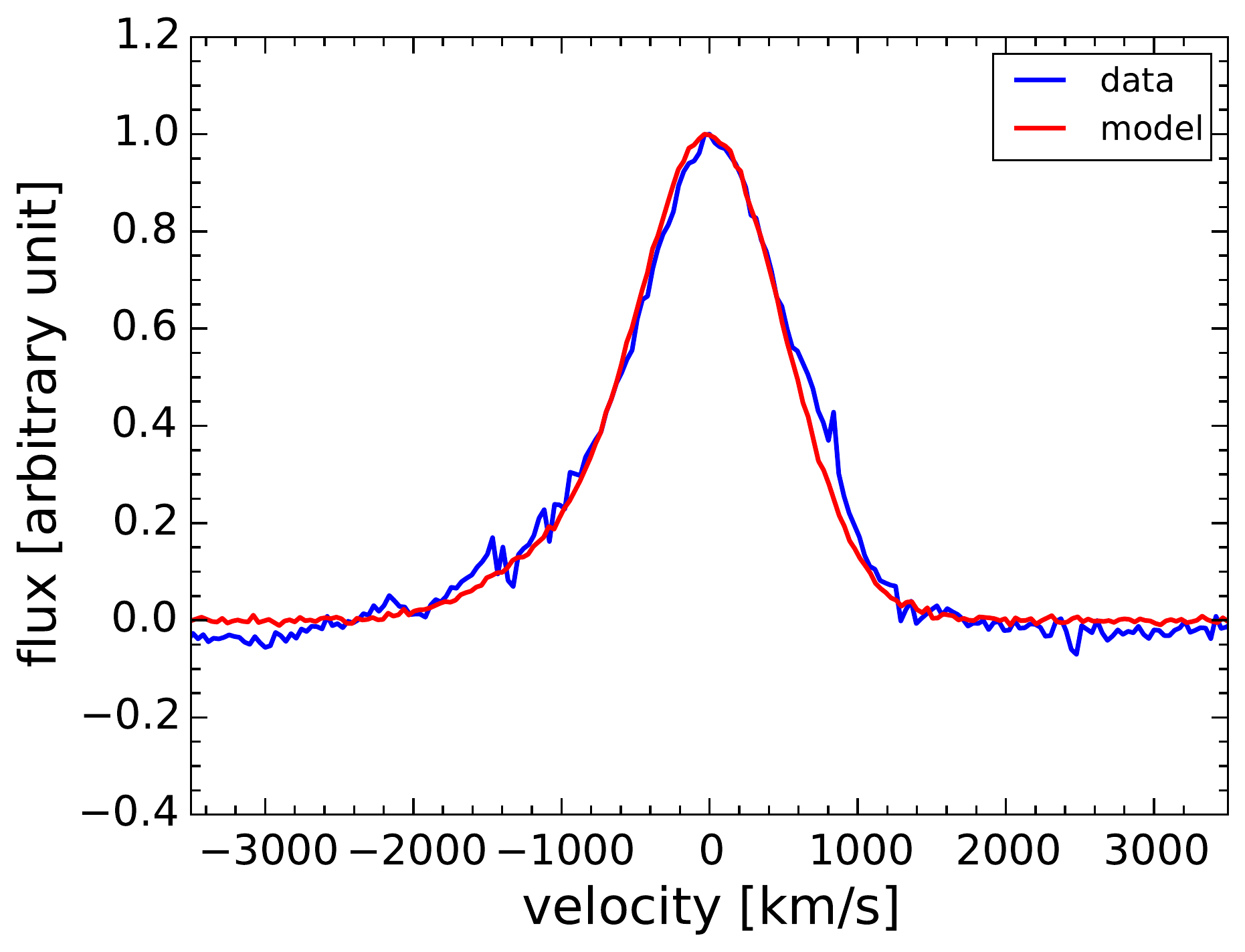}
 
 \caption{[OIII]$\lambda5007$ emission lines extracted from a nuclear region of 0.25\arcsec$\times$ 0.25\arcsec\ . The blue line is the \oiii\ emission line of LBQS0109  after subtracting all the other best-fit components. The red line is the ionised emission line obtained from the simulated data.  }   
\label{fig:model_spectrum}
   \end{figure}

\begin{figure}
   \centering
\includegraphics[width =0.5\textwidth]{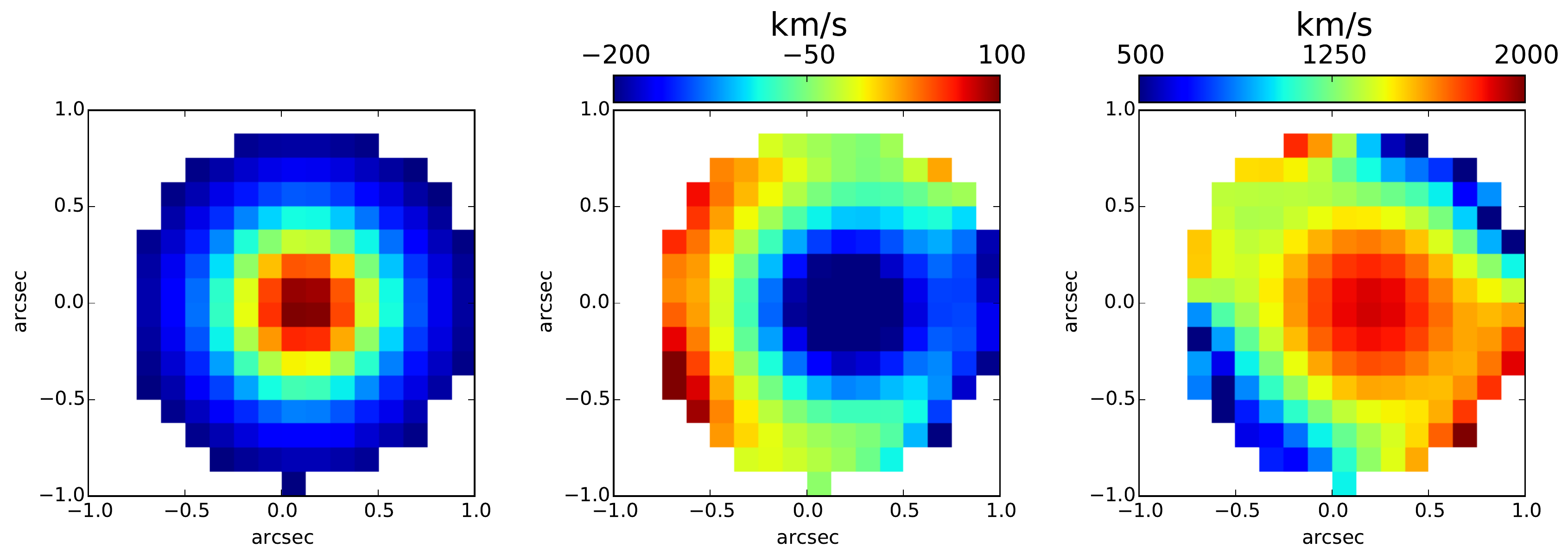}
 
 \caption{[OIII] flux, velocity and velocity dispersion map obtained from the simulated \oiii\ data. The regions with S/N $<2$ are masked out.  }

\label{fig:model_velocity_map}
   \end{figure}

The comparison between the \oiii\ line profile extracted from the simulated data and that from the observations is  shown in Figure \ref{fig:model_spectrum}. Both spectra are extracted from a nuclear region 
of 0.25\arcsec$\times$ 0.25\arcsec. The  simulated emission line presents a prominent blue wings similarly to the real spectrum of LBQS0109. We then performed the kinematic analysis on the simulated data as described in Section \ref{sec:fitting}. The flux, velocity and velocity dispersion maps (Fig. \ref{fig:model_velocity_map}) obtained from the simulations are similar to those shown in Figure \ref{fig:velocity_maps} extracted from observations. Indeed, the velocity map shows  blue shifted velocities in the outflow region extended over 0.5 \arc (i.e $\sim$4.2 kpc). So, the simulation confirms that the blue-shifted region suffers from beam smearing and its size does not match the real radius of the outflow. 

\begin{figure}
   \centering
\includegraphics[width =0.5\textwidth]{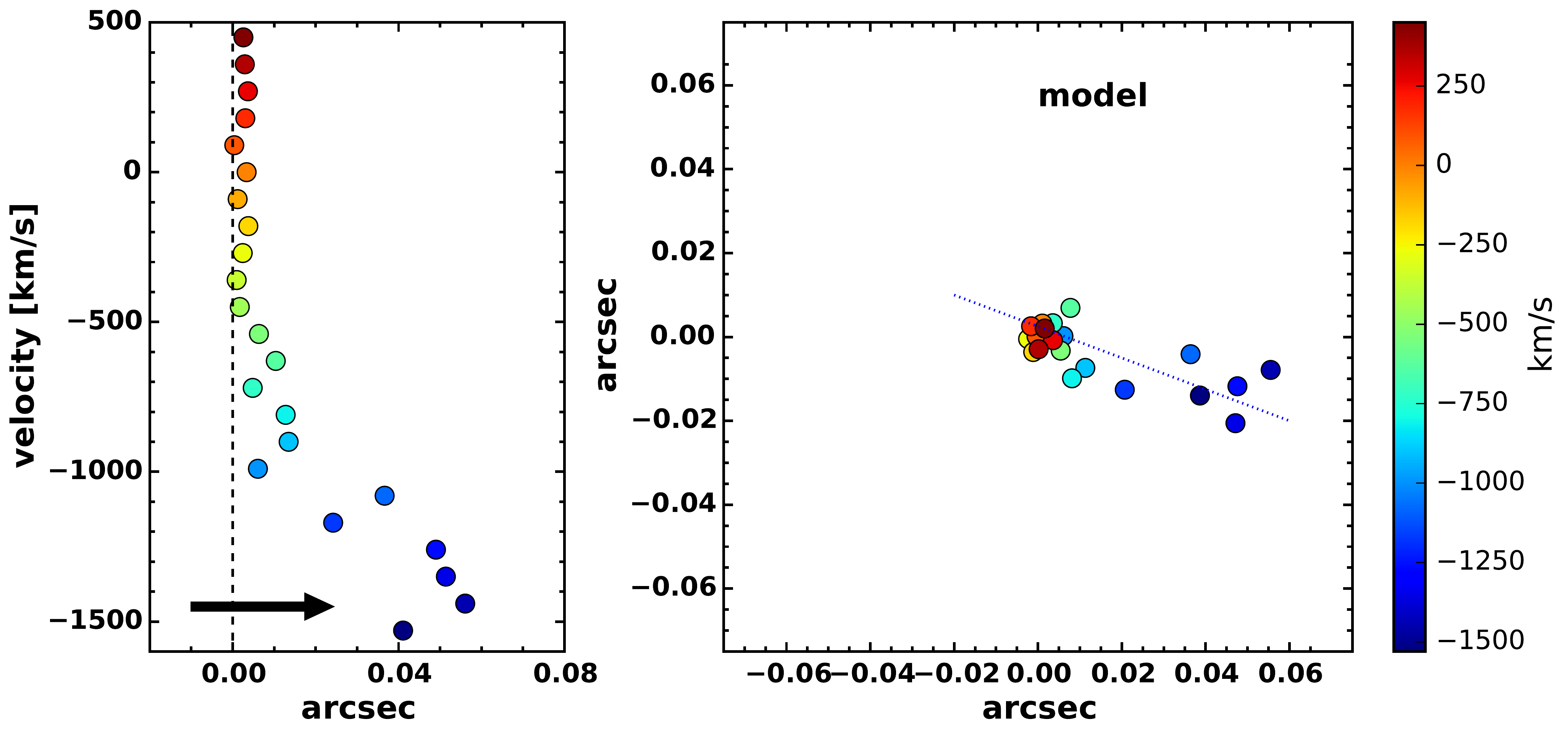}
 
 \caption{Spectroastrometry results obtained from the simulated data. On the left,  centroid offset of \oiii\ emission respect to continuum emission (dashed line) at different wavelength. On the right, the photo-centroid position in each velocity. Colour coding corresponds to velocity offset.}   
\label{fig:model_spectroastrometry}
   \end{figure}

Since the simulated outflow well describes the asymmetric  \oiii\  profile and the velocity gradient observed in our datasets, we carried out the spectroastrometry analysis   on the simulated data to estimate the average velocity and radius of the ionised gas in order  to compare these derived values with the input parameters. 
The offset of the photo-centroids at different velocities are shown in Figure \ref{fig:model_spectroastrometry}. The radius $R_o$ ($\sim 0.05$ \arc) and the average velocity $v(R_o)$ ($\sim 1450 $ km/s) estimated  from spectroastrometry are consistent to the initial input parameters. So, provided that the S/N is high enough, the spectroastrometry method can
provide the position and the velocity of outflows with an accuracy well below the seeing limit. Moreover, since the integral flux of \oiii\ at high blue-shifted velocity ($ -2500 {\rm km/s} < v <  v(R_o)$) is consistent with about half of the total flux emitted from the ionised outflowing gas, the outflow luminosity can be inferred from:
\begin{equation}
L_{o} = 2\int_{-\infty}^{\lambda_0} L_\lambda(\lambda)d{\lambda}
\label{eq:lum}
\end{equation}
where $\lambda_0$ is the wavelength corresponding to $v_0$.
In the cases where the \oiii\ emission line was well described by a multi-Gaussian fit,  the luminosity value calculated with equation (\ref{eq:lum}) and the one estimated from the broad Gaussian component, as typically done in the literature, are consistent within the errors. So, in the multi-Gaussian fit we estimated the $L_o$ from the broad component and in the other case we measured the luminosity of the gas using the equation (\ref{eq:lum}).

Since the narrow \hb\ component has the same profile of [OIII], we can measure the outflow luminosity for both the emission lines using the same equation. From the luminosity we can get the mass of the ionised outflows by using equation (\ref{eq:mass}) and (\ref{eq:masshb}). 
As the \oiii/\hb\ ratio measured in each QSO is $\sim10$, the mass inferred by \hb\ is $\sim2$ times larger than that estimated by \oiii\ (see eq. \ref{eq:ratio}).  
The luminosity and mass of the outflow are listed in Table \ref{tab:summary} and \ref{tab:outflow}, respectively.
Since HE0251 is not spatially resolved, we cannot estimate the outflow luminosity using the method described above. In this case we fitted the \oiii\ with  two Gaussian components  and we inferred the outflow luminosity from the broad Gaussian component.

\section{Results}\label{sec:results}

In order to investigate the nature of the ionised wind, we need to estimate the main quantities of the outflows: mass outflow  rate, momentum rate and kinetic power. Theoretical models (e.g  \citealt{Zubovas:2012}, \citealt{Faucher-Giguere:2012}) predict tight relations between these quantities and the AGN bolometric luminosity, \lagn. In particular the correlation between the momentum rate and \lagn\  provides an indicator of the nature of the feedback mechanism. In this section  we derive the main properties of the AGNs in our sample and  compare them with the prediction of the  models.
As  explained in Section \ref{sec:simulation}, the low spatial resolution of our observations does not allow us to discern whether our outflows are single explosive events or refilled with clouds ejected from the galactic disk. Hence,  we assume a simple model  where a single ionised cloud is ejected outward 
of the nuclear region and the mass outflow rate is equal to the mass  of the outflow divided by the dynamical timescale (see eq. 1).
The dynamic time is the time that a clump of ionised gas in outflow takes to reach a fixed distance from the QSO.   
We derive   \mion\ values in the range $6 - 700$ \sfr.  We assume an average outflow mass rate error of $\pm50\%$, which takes in account the uncertainties associated with the gas physical properties (i.e. density and temperature), flux calibrations and $R_0$ and $v_0$ estimates due to projection effects.   
For each source we estimate two value of \mion\ that are calculated by using \hb\ and \oiii\  respectively. 
 Since the \oiii\ outflow luminosities are $\sim 10$ times larger than the ones from \hb, the outflow masses inferred from the hydrogen emission line are $\sim$2 times larger than those estimated by \oiii. 
\begin{figure*}
\centering
\includegraphics[width = .47\textwidth]{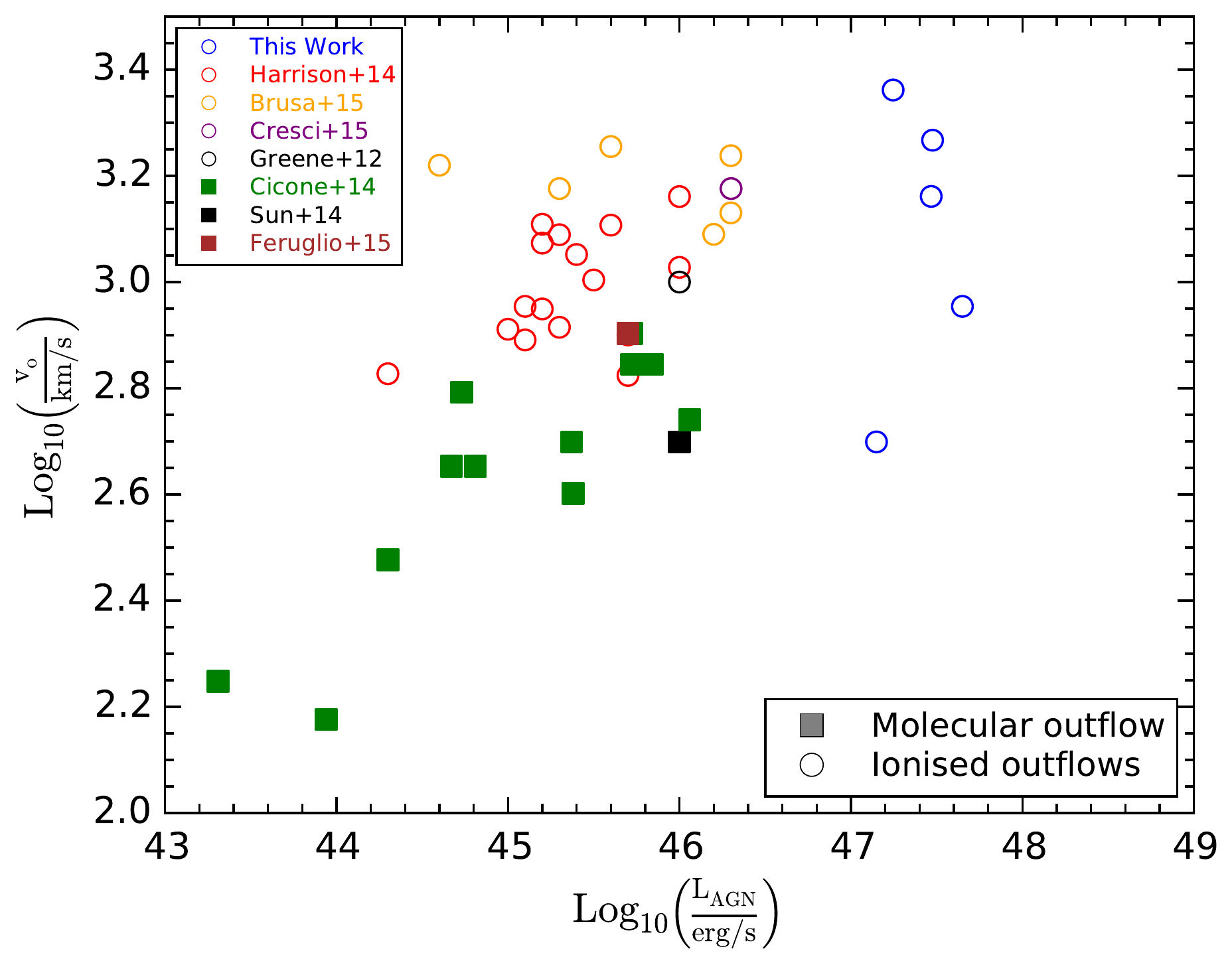}
\includegraphics[width = .49\textwidth]{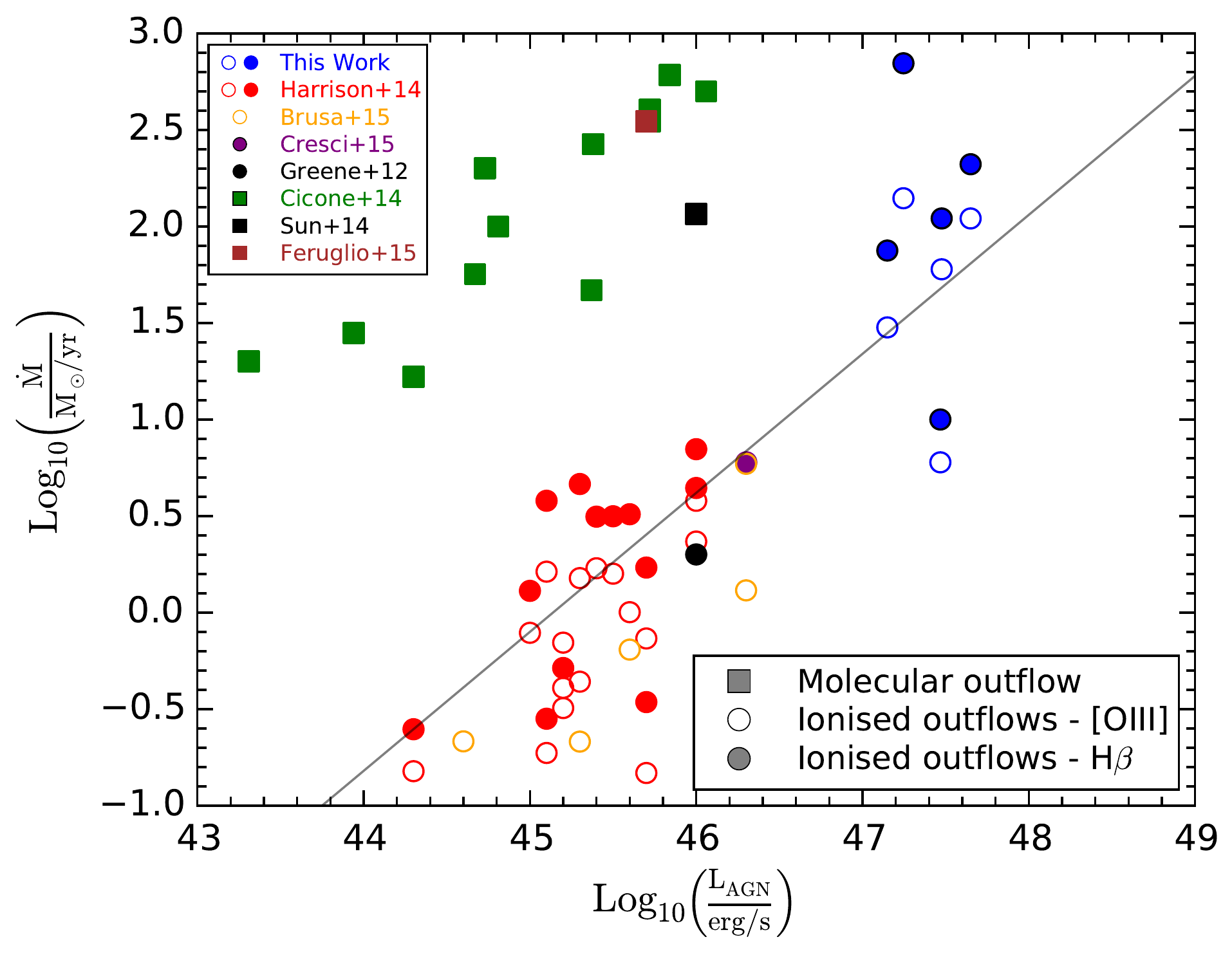}

\caption{ Left : Outflow velocity as a function of the AGN bolometric luminosity. The blue circles denote the results from this work; the open circles mark the velocities of ionised gas (mainly \oiii): the red, orange, purple and black circles are the estimates obtained from  \cite{Harrison:2014}, \cite{Brusa:2015}, \cite{Cresci:2015} and \cite{Greene:2012}, respectively.  The green, black and brown squares denote the velocities of the molecular outflows  from \cite{Cicone:2014}, \cite{Sun:2014} and \cite{Feruglio:2015}. 
Right: Outflow rates as a function of the AGN bolometric luminosity.  Notation is  the same of the left panel expect that 
 open circles mark the estimates obtained with [OIII]-inferred masses, while the filled ones denote the estimates based on \hb. 
We recalculated the outflow properties inferred by previous works  to make them consistent with our estimates (see text).
The solid line is the best fit relation to the averages of filled and empty circles. 
}
\label{fig:plot_outflow_rate}
 \end{figure*}

In Figure~\ref{fig:plot_outflow_rate} we plot the outflow velocity and mass  rate as a function of the AGN luminosity.
\lagn\ is derived by using the relation  \lagn \ $\sim 6\lambda L(\lambda5100\AA)$  from \cite{Marconi:2004}. The solid and hollow blue points are the \mion\ estimated for five out of six QSOs; we could not infer the value of HE0251 since we were not able to measure the size of the outflow. 
The red points represent the ionised outflows observed in type 2 AGN at redshift $0.08 \lesssim z \lesssim 0.2$  \citep{Harrison:2014}; the orange and purple points are from \cite{Brusa:2015} and  \cite{Cresci:2015} who mapped  ionised outflows in 6 X-ray selected, obscured QSOs at $z\sim1.5$; the black circle corresponds to the the ionised outflow in a obscured radio-quiet QSO at $z\sim$0.123  \citep{Greene:2012}.
For consistency with our work we re-calculated the outflows properties using an electronic density of $n_e$ =500 cm$^{-3}$ and a temperature of $T_e = 10^4 $ K.
Note that the outflow masses from literature may  be overestimated since  we used the luminosities  from the total \oiii\ and \hb\ profiles.


In addition to ionised outflows, we compare our results with those obtained from molecular outflows ( green, black and brown solid square; \citealt{Cicone:2014}, \citealt{Sun:2014} and \citealt{Feruglio:2015}, respectively). The molecular outflow properties were re-estimated assuming a  shell-like cloud model (eq. \ref{eq:rate}). 

 In  Figure  \ref{fig:plot_outflow_rate} we show that  the outflow velocity $v_0$  and outflow rate are correlated   with the AGN luminosity, although with a large scatter. 
The increase in velocity and outflow rate with increasing AGN luminosity is consistent with the idea that a luminous  AGN pushes away the surrounding gas through  a radiatively driven fast wind whose kinetic power is a fraction of the AGN luminosity.   However, it is not easy to establish a direct  relation between AGN luminosity, velocity and outflow rate. The acceleration process of the outflow  by the fast wind and the fraction of the kinetic power injected, might vary from object to object; moreover,  the observed AGN luminosity may not represent the long-term average luminosity  which is at the end responsible for driving the outflow. For these reasons we do not expect  tight correlations between velocity, outflow rate and AGN luminosity, as observed.   
Indeed, previous studies on more heterogenous samples, but smaller luminosity ranges, did not find any significant correlation between the outflow velocities and the AGN luminosities (e.g. \citealt{Veilleux:2013}, \citealt{Brusa:2015}). 
In the luminosity range where we have data from both CO and [OIII] ($10^{44.5}  \ {\rm erg/s} \ \lesssim {\rm L_{\rm AGN}} \lesssim 10^{46.5} \ {\rm erg/s} $), the velocities of the ionised outflows are a factor $\sim$2 larger than those of  the molecular gas, consistently with larger masses of molecular gas in  both momentum- and energy-driven scenarios, but also with different acceleration/deceleration processes.     
However,  given the heterogeneous nature of the sample and the  non-uniform measurements of outflow velocities we cannot draw any firm conclusions.
Moreover, there is only one galaxy (black circle and square in Fig. \ref{fig:plot_outflow_rate}) where both molecular and ionised outflows are detected.  
This discrepancy is even more evident when comparing molecular with ionised outflow rates.
\cite{Cicone:2014}  fit a  a log-linear relation  between $\dot{M}$ and \lagn\ for their molecular outflows finding $Log_{10}(\dot{M}) = 2.84+0.720\times Log_{10}(L_{AGN}/10^{46} {\rm erg/s} )$ where we have corrected for the factor 3 discrepancy described in Section \ref{sec:spectra}.
Here, we fit the same relation between $\dot{M}$ and \lagn\ fixing the  slope to that of \cite{Cicone:2014}:
\begin{equation}
Log_{10}(\dot{M}) = (0.6\pm0.2)+0.720\times Log_{10}(L_{AGN}/10^{46} {\rm erg/s})
\end{equation}
Since the velocities and the radii of ionised outflows are similar to those of molecular ones, we interpret the offset between the two \lagn - ${\dot M }$ correlations  as an indication that   the ionised gas only traces a small fraction of the total gas mass.  Comparing the normalization of the \lagn - ${\dot M }$ relation inferred from our data with that obtained from molecular outflows, we  can infer  that the outflow rate of the ionised gas is a small fraction ($< 10\%$) of the molecular one, suggesting that outflow masses are dominated by molecular gas.  However, since  molecular and ionised outflow rates are not estimated in the same objects and the relations have quite a large scatter, it is not possible to estimate the exact fraction of ionised gas in the outflows.  
On the other hand, the different normalisation could indicate that two different acceleration mechanisms are at work. 
 Future ALMA observations of the molecular gas in these objects are essential  to distinguish between these two different scenarios.



\begin{figure}
\centering

\includegraphics[width = .5\textwidth]{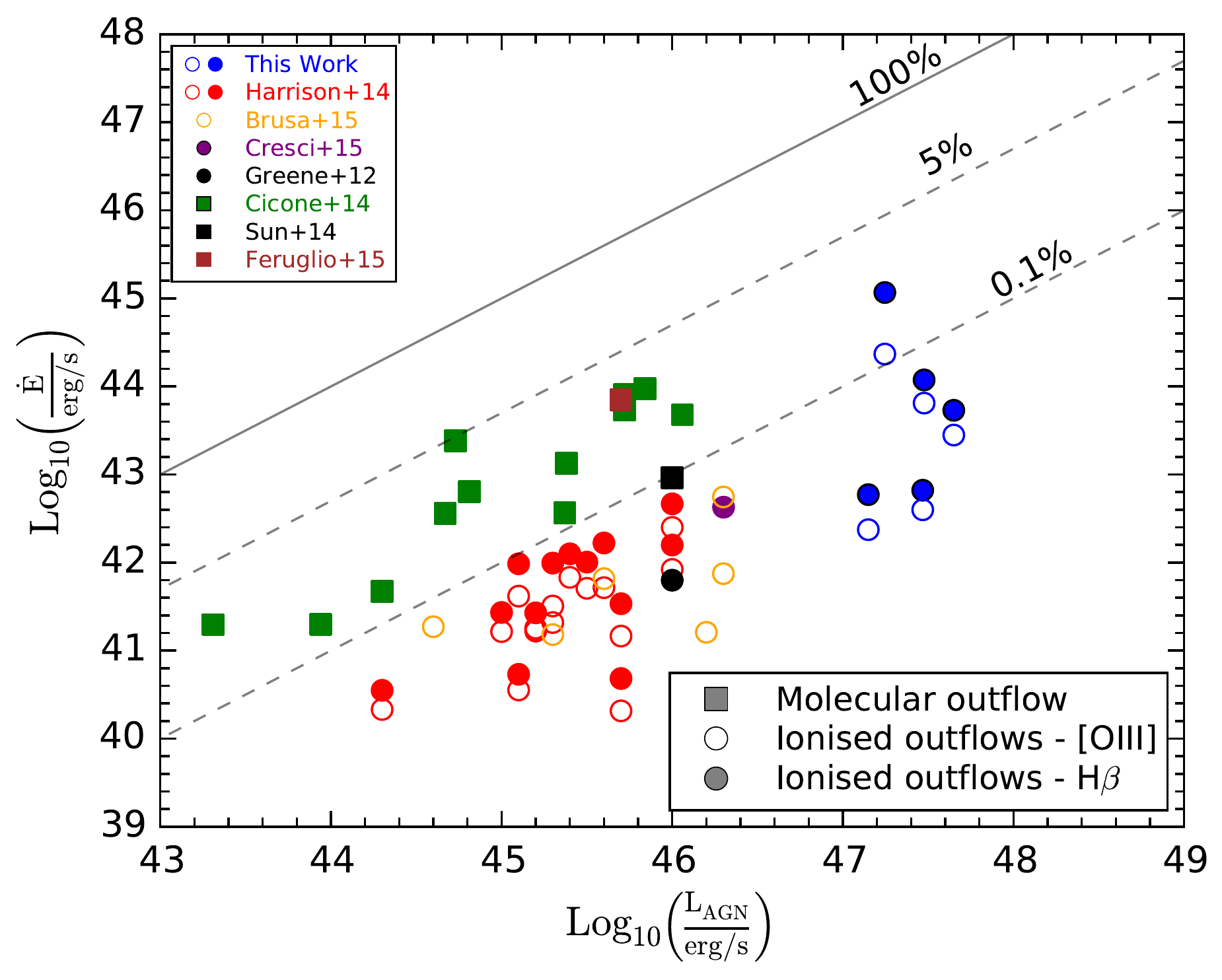}

\caption{Kinetic power as a function of the AGN bolometric luminosity. Symbols and colours as in fig. \ref{fig:plot_outflow_rate}. The solid, dashed and dotted line correspond to $P_k = 100\%,5\%,0.1\% L_{AGN}$ respectively. }
\label{fig:plot_power}
 \end{figure}

Figure \ref{fig:plot_power} shows the kinetic power of  outflowing gas as a function of AGN luminosity where the outflow kinetic energy rate is given by

$$
\dot E_{\rm o} = \frac{1}{2}\dot M_{o} v_o^2 = \frac{1}{2}\frac{M_{o}}{R_o} v_o^3
$$
We indicate using solid, dashed and dotted lines the locus of points having an outflow kinetic power that is 100\%, 5\% and 0.1\% of the AGN luminosity, respectively.
The recent AGN feedback models   (e.g. \citealt{King:2010}; \  \citealt{Zubovas:2012}; \citealt{Lapi:2014}) predict a coupling efficiency between AGN-driven outflows and AGN power of about $\sim$5\% which is needed to  explain the $M_{BH}$ - $\sigma$ relation observed in local galaxies.
Although the molecular outflow observations are consistent with the models within the error, the kinematic power estimated from  ionised outflows is  only $<0.1\%$ of the \lagn. This percentage is too low to explain  the $M_{BH}$ - $\sigma$ relation. Since the kinetic energy is proportional to the outflow mass, we cannot discern whether  the nature of ionised outflows is different with respect to molecular ones or, as discussed above, the \oiii\ (and \hb) line emission traces only a small fraction ($< 10\%$) of the total outflowing gas.

\begin{figure}
\centering

\includegraphics[width = .5\textwidth]{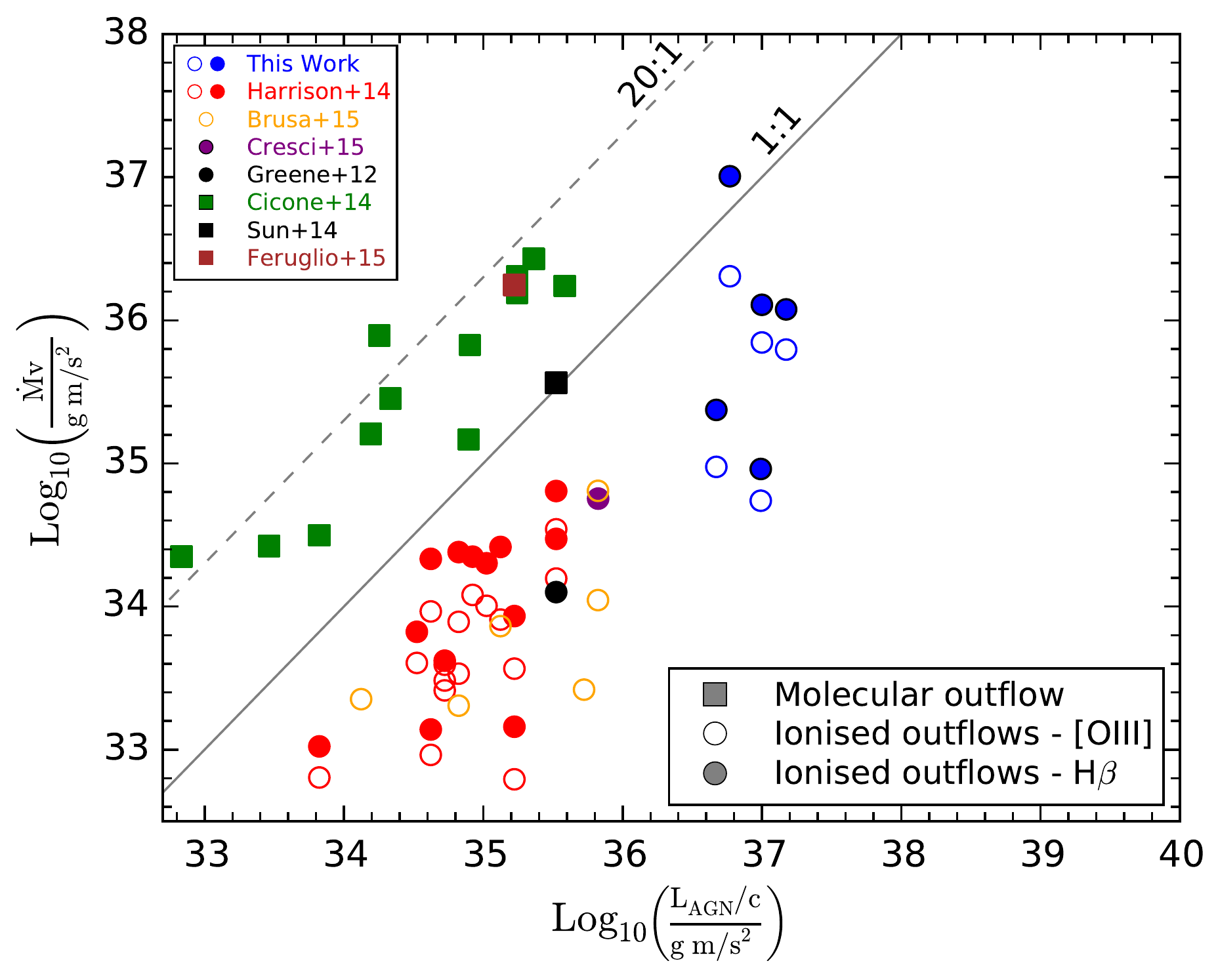}

\caption{Outflow momentum rate as a function of photon momentum of the AGN . Symbols and colours as in Fig. \ref{fig:plot_outflow_rate}. The dotted and dashed  line correspond to \mom$\sim20$\lagn / c and \mom$\sim$\lagn / c respectively.
}
\label{fig:plot_momentum}
 \end{figure}

The last fundamental parameter of the outflows is the outflow momentum rate defined as $v_o \dot M_{o}$. 
Energy-driven outflow models predict that the momentum of the large-scale  outflow is boosted compared to the nuclear wind (or AGN radiation pressure momentum ), i.e. $v_o \dot M_{o}\sim20$\lagn / c \citep{Zubovas:2012}. In Figure \ref{fig:plot_momentum} we observe that molecular outflows follow the relation predicted by models (see \citealt{Cicone:2014} for more details) while ionised outflows are below the 1:1 relation.
This could simply be the consequence of the discrepancy between the outflow masses of ionised and molecular gas.

Finally, we may conclude  that the main difference between molecular and ionised outflows is the fraction of total gas mass pushed away from the AGN-driven wind.  This is confirmed by the larger  outflow rates (right panel of Fig.~  \ref{fig:plot_outflow_rate}) and kinetic power (Fig.  \ref{fig:plot_power}) of molecular outflows, and by  their smaller outflow velocities (left panel of Fig. \ref{fig:plot_outflow_rate}). 


\section{Conclusions}\label{sec:conclusions}
We performed seeing-limited, near-IR integral-field spectroscopic observations with SINFONI of a sample of six, high luminosity (${\rm L_{\rm bol}} = 10^{13}-10^{14}$ \lsun) QSOs at redshift $z\sim2.4$. \oiii\ emission lines, redshifted into the H-band, are characterised by large FWHMs ($>1000$ km/s) and prominent blue wings, indicative of fast outflows accelerated by the powerful AGN. We find that:
\begin{itemize}

\item The \oiii\ emission line is spatially resolved in five out of six sources and extended over several kiloparsecs.

\item The analysis of the \oiii\ kinematical maps suggests the presence of conical outflows associated with regions of high velocity dispersions ( $>$ 500 km/s) 
The physical properties of the outflows, i.e. mass, outflow rate, kinetic energy and momentum rate, have been estimated with a new method based on spectroastrometry which is not affected by PSF smearing, at variance with results from previous works. The reliability of this method has been confirmed with a simple simulation but  more accurate modelling will be developed  in future work. 


\item  Both ionised and molecular outflow velocities are weakly correlated with the observed AGN luminosity, despite the large scatter, and at a given AGN luminosity the velocity of the ionised gas is  roughly a factor 2 larger than that of the molecular one. 
However, given the heterogenous nature of the combined sample we cannot draw any firm conclusions. 


\item Mass and momentum rates, as well as kinetic powers, increase with AGN bolometric luminosity in a similar way to what is observed in molecular outflows in the local universe. The ionised gas properties define relations with AGN luminosity which are parallel with those of molecular gas. In particular, ionised outflow rates are $\sim 50$ times lower than molecular ones. The  kinetic power carried by ionised outflows is of the order of $\sim0.1-0.05\%$ of the AGN luminosity compared to $\sim5\%$ for molecular outflows.  Finally, momentum rates are of the order of \lagn/c, a factor $\sim20-50$ smaller than for molecular outflows.

\item These discrepancies between ionised and molecular outflows can be explained  with the fact that ionised gas traces a smaller fraction of the total gas mass.  
Alternatively, they are the indication of different acceleration mechanisms for the molecular and the ionised gas. 
Observations with ALMA would allow us to measure the molecular gas mass in these objects therefore distinguishing between these two scenarios.

%

%

\end{itemize}

\begin{acknowledgements}
We thank the anonymous referee for comments and suggestions that improved the paper. We acknowledge financial support from INAF under the contracts PRIN-INAF-2011 (``Black Hole growth and AGN feedback through cosmic time") and PRIN MIUR 2010-2011 (``The dark Universe and the cosmic evolution of baryons"). 
MB acknowledges support from the FP7 Career Integration Grant ``eEASy'' (CIG 321913).
CF gratefully acknowledges financial support from PRIN MIUR 2010-2011, project ``The Chemical and Dynamical Evolution of the Milky Way and Local Group Galaxies", prot. 2010LY5N2T. EP acknowledges financial support from INAF under the contract PRIN-INAF-2012. Funding for this work has also been provided by the Israel Science Foundation
grant 284/13.

\end{acknowledgements}

\bibliographystyle{aa} 
\bibliography{bibliography}

\end{document}